\documentclass[aps,prb,longbibliography,twocolumn,superscriptaddress,floatfix,10pt]{revtex4-1}
\usepackage[utf8]{inputenc}

\usepackage{natbib}
\usepackage{graphicx}
\usepackage{latexsym}
\usepackage{amssymb}
\usepackage{amsmath}
\usepackage{amsfonts}
\usepackage{gensymb}
\usepackage{bm}
\usepackage{hyperref}
\usepackage{braket}
\usepackage{tikz}
\usepackage{rotating}
\usetikzlibrary{arrows}
\usetikzlibrary{decorations.pathreplacing,decorations.markings}
\usepackage[normalem]{ulem} 

\newcommand{\refcite}[1]{Ref.\,\onlinecite{#1}}
\newcommand{\eqnref}[1]{Eq.\,(\ref{#1})}
\newcommand{\figref}[1]{Fig.\,\ref{#1}}
\newcommand{\sfigref}[2]{Fig.\,\hyperref[#1]{\ref{#1}(#2)}}

\newcommand{\secref}[1]{Sec.\,\ref{#1}}
\newcommand{\appref}[1]{Appendix~\ref{#1}}

\newcommand{\nn}{\nonumber}

\hypersetup{
  colorlinks = true,
  urlcolor = blue,
  pdfauthor = {Wilbur Shirley, Kevin Slagle, Zhenghan Wang, Xie Chen},
  pdftitle = {Shirley et al. - 2017 - Fracton Models on General Three-Dimensional Manifolds}
}

\begin{document}

\title{Fracton Models on General Three-Dimensional Manifolds}
\date{\today}

\author{Wilbur Shirley}
\affiliation{Department of Physics and Institute for Quantum Information and Matter, California Institute of Technology, Pasadena, California 91125, USA}

\author{Kevin Slagle}
\affiliation{Department of Physics, University of Toronto, Toronto, Ontario M5S 1A7, Canada}

\author{Zhenghan Wang}
\affiliation{Microsoft Station Q and Department of Mathematics, University of California, Santa Barbara, CA 93106-6105 USA}

\author{Xie Chen}
\affiliation{Department of Physics and Institute for Quantum Information and Matter, California Institute of Technology, Pasadena, California 91125, USA}

\begin{abstract}
Fracton models, a collection of exotic gapped lattice Hamiltonians recently discovered in three spatial dimensions, contain some `topological’ features: they support fractional bulk excitations (dubbed fractons), and a ground state degeneracy that is robust to local perturbations. However, because previous fracton models have only been defined and analyzed on a cubic lattice with periodic boundary conditions, it is unclear to what extent a notion of topology is applicable. In this paper, we demonstrate that the $X$-cube model, a prototypical type-I fracton model, can be defined on general three-dimensional manifolds. Our construction revolves around the notion of a singular compact total foliation of the spatial manifold, which constructs a lattice from intersecting stacks of parallel surfaces called leaves. We find that the ground state degeneracy depends on the topology of the leaves and the pattern of leaf intersections. We further show that such a dependence can be understood from a renormalization group transformation for the X-cube model, wherein the system size can be changed by adding or removing 2D layers of topological states. Our results lead to an improved definition of fracton phase and bring to the fore the topological nature of fracton orders.

\end{abstract}

\maketitle

\section{Introduction}

Characterization and classification of quantum phases of matter is a fundamental problem of physics.  Spectacular progress has been made in the last decade for topological phases of matter, especially those with short-range entanglement and with long-range entanglement in two spatial dimensions.
Topological quantum field theory (TQFT) is the framework in which the regnant theories of these topological phases are formulated. Recently, an intriguing class of gapped Hamiltonians, referred to as fracton models in this paper, have been proposed as potential new topological phases of matter. \cite{VijayFracton,YoshidaFractal,Sagar16,HaahCode,ChamonModel,ChamonModel2,VijayCL, MaLayers,HsiehPartons,HalaszSpinChains,VijayNonabelian,Petrova_Regnault_2017}
These models appear in three spatial dimensions and have ground states exhibiting long-range entanglement. Certain topological features \cite{Schmitz_2017,BernevigEntropy,HermeleEntropy,DevakulCorrelationFunc,ShiEntropy,WilliamsonUngauging,PremGlassy} such as robust ground state degeneracy and fractional excitations rear their heads in fracton models. But these models clearly do not fit into the standard TQFT framework since their ground state degeneracies (GSD) are not topologically invariant, which is a salient feature of current TQFTs. In this paper, we investigate the nature of the underlying physics in these fracton models: topological, geometric, or something yet else?

Quantum field theory (QFT) provides powerful descriptions of many-body quantum physics. Phases of matter with intrinsic topological order can be characterized either by the existence of a low energy TQFT limit or by an anyon model that captures the algebraic structure of fractional excitations in the plane. We are thus motivated to ask whether fracton models have low energy descriptions that resemble TQFTs. (A QFT description of the X-cube model with a cut-off is proposed in \refcite{Slagle17}.)

A TQFT assigns a Hilbert space $V(Y)$ to each spatial manifold $Y$ satisfying some formal properties of a QFT;
the Hilbert space $V(Y)$ is the ground state manifold.  An intrinsic topological order manifests itself in the robust ground state degeneracy $V(Y)$ that depends only on the topology of the spatial manifold $Y$.  Fracton models do not fit into this framework because their GSD is not solely determined by the topology of the spatial manifold $Y$. Thus, we are interested in a modification such that the ground state manifold $V(Y,s)$ will depend not only on the topology of $Y$ but also some extra structure $s$ of $Y$ analogous to a $G$-bundle in gauge theory or a spin structure for fermions.  A condensed matter system on a closed (compact without boundary) spatial manifold $Y$ is defined through a Hamiltonian on a lattice $\Delta$ in $Y$, which is a cellulation of $Y$ mathematically.  In traditional topological phases, the ground state manifold $V(Y,\Delta)$ is independent of the lattice $\Delta$, while in fracton models, $V(Y,\Delta)$ depends on the lattice $\Delta$ in intricate ways.  One hope is that for a judiciously chosen sequence of lattices $\Delta_i$, the ground state manifolds $V(Y,\Delta_i)$ converge to a well-defined limit $V(Y,s)$, presumably infinitely dimensional, and their dependence on lattices $\Delta_i$ reduces to the well-defined structure $s$ of $Y$.  Then according to the structure $s$ being regarded as topological, geometric, or something yet else, we will classify the fracton models as phases possessing such a character.  

In this paper, a first step is made towards such a generalized TQFT for the $X$-cube model. As explained in the concluding section, we conjecture that the extra structure is a singular compact total foliation, inspired by the notion of total foliation of a $3$-manifold.\cite{Hardorp_1980} Since a foliation of a $3$-manifold is regarded as a topological structure, we suggest that the $X$-cube model can be considered to be a new kind of generalized topological order.

In particular, we show that the $X$-cube model,\cite{Sagar16} originally defined on the three-dimensional torus, can be defined on other closed 3-manifolds as well. To do so, we employ a singular compact total foliation of a 3-manifold, which partitions the manifold into three sets of transversely intersecting  parallel surfaces in the complement of a (possibly empty) singular subset consisting of singular leaves. The ground state degeneracy (and its size-independent correction) depends on the topology of these leaves and the intersections of the leaves. We show that the relation between the ground state degeneracy and the foliation can be easily understood in terms of an entanglement renormalization group transformation of the $X$-cube model where system size can be increased or decreased by adding or removing 2D layers of toric code topological states. 

The total foliation structure is well-suited for the potential construction of a continuous limit description of the phase. Additionally, we give a spherical leaf construction of the $X$-cube model for any 3-manifold. Remarkably, we find that 
the $X$-cube model in the spherical leaf construction reduces to the 3D toric code model with traditional topological order of a discrete vector gauge theory.

The paper is structured as follows: In \secref{sec:review}, we review the $X$-cube model on the $3$-torus $T^3$. In \secref{sec:lattices}-\ref{sec:manifolds}, we extend the $X$-cube model to other closed 3-manifolds: the spherical leaf construction in \secref{sec:lattices} and the singular compact total foliation construction in \secref{sec:manifolds}.  In \secref{sec:RG}, we present the entanglement renormalization procedure for the $X$-cube model. In \secref{sec:ZN}, we show that these results can be generalized to the $\mathbb{Z}_N$ version of the X-cube model.


\section{Review of X-cube model on three-torus}
\label{sec:review}

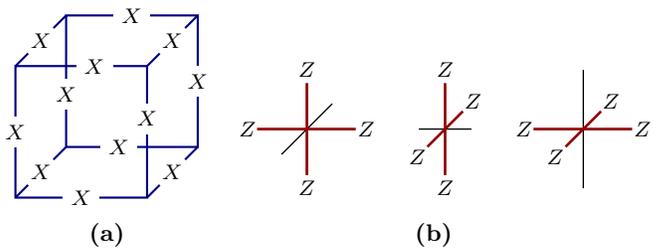
\begin{figure}
    \begin{tikzpicture}
        \pgfmathsetmacro{\l}{1.75}

        \draw[line width=.8,color={rgb:black,1;blue,1}] (0,0,0) -- ++(-\l,0,0) -- ++(0,-\l,0) -- ++(\l,0,0) -- cycle;
        \draw[line width=.8,color={rgb:black,1;blue,1}] (0,0,-\l) -- ++(-\l,0,0) -- ++(0,-\l,0) -- ++(\l,0,0) -- cycle;
        
        \draw[line width=.8,color={rgb:black,1;blue,1}] (0,0,0) -- ++(0,0,-\l*.25);
        \draw[line width=.8,color={rgb:black,1;blue,1}] (0,0,-\l) -- ++(0,0,\l*.25);
        \draw[line width=.8,color={rgb:black,1;blue,1}] (0,-\l,0) -- ++(0,0,-\l*.25);
        \draw[line width=.8,color={rgb:black,1;blue,1}] (0,-\l,-\l) -- ++(0,0,\l*.25);
        \draw[line width=.8,color={rgb:black,1;blue,1}] (-\l,0,0) -- ++(0,0,-\l*.25);
        \draw[line width=.8,color={rgb:black,1;blue,1}] (-\l,0,-\l) -- ++(0,0,\l*.25);
        \draw[line width=.8,color={rgb:black,1;blue,1}] (-\l,-\l,0) -- ++(0,0,-\l*.25);
        \draw[line width=.8,color={rgb:black,1;blue,1}] (-\l,-\l,-\l) -- ++(0,0,\l*.25);
        
        \pgfmathsetmacro{\ri}{1}
        \pgfmathsetmacro{\di}{.3}
        \draw[line width=.5] (.45*\l+\ri,-\l/2-\di,-\l) -- ++(0,0,\l);
        \draw[line width=1,color={rgb:black,.75;red,1}] (\ri+.125,-\l/2-\di,-\l/2) -- ++(\l*.75,0,0);
        \draw[line width=1,color={rgb:black,.75;red,1}] (.45*\l+\ri,-\l*.85-\di,-\l/2) -- (.45*\l+\ri,-\l*.15-\di,-\l/2);

        \pgfmathsetmacro{\rii}{\ri+1.3*\l}
        \draw[line width=.5] (\rii,-\l/2-\di,-\l/2) -- ++(\l*.4,0,0);
        \draw[line width=1,color={rgb:black,.75;red,1}] (.2*\l+\rii,-\l/2-\di,-\l*.85) -- (.2*\l+\rii,-\l/2-\di,-\l*.15);
        \draw[line width=1,color={rgb:black,.75;red,1}] (.2*\l+\rii,-\l*.85-\di,-\l/2) -- (.2*\l+\rii,-\l*.15-\di,-\l/2);
        
        \pgfmathsetmacro{\riii}{\rii+.8*\l}
        \draw[line width=.5] (.45*\l+\riii,-\l*.95-\di,-\l/2) -- ++(0,.9*\l,0);
        \draw[line width=1,color={rgb:black,.75;red,1}] (.45*\l+\riii,-\l/2-\di,-\l*.85) -- (.45*\l+\riii,-\l/2-\di,-\l*.15);
        \draw[line width=1,color={rgb:black,.75;red,1}] (\riii+.125,-\l/2-\di,-\l/2) -- ++(\l*.75,0,0);
        
        \draw (0,0,-\l/2) node[fill=none] {\footnotesize $X$};
        \draw (0,-\l,-\l/2) node[fill=none] {\footnotesize$X$};
        \draw (-\l,0,-\l/2) node[fill=none] {\footnotesize$X$};
        \draw (-\l,-\l,-\l/2) node[fill=none] {\footnotesize$X$};
        \draw (0,-\l*.4,0) node[fill=white] {\footnotesize$X$};
        \draw (0,-\l/2,-\l) node[fill=white] {\footnotesize$X$};
        \draw (-\l,-\l*.6,-\l) node[fill=white] {\footnotesize$X$};
        \draw (-\l,-\l/2,0) node[fill=white] {\footnotesize$X$};
        \draw (-\l*.4,0,0) node[fill=white] {\footnotesize$X$};
        \draw (-\l/2,-\l,0) node[fill=white] {\footnotesize$X$};
        \draw (-\l/2,0,-\l) node[fill=white] {\footnotesize$X$};
        \draw (-\l*.6,-\l,-\l) node[fill=white] {\footnotesize$X$};
        
        \draw (\ri,-\l/2-\di,-\l/2) node[fill=none] {\footnotesize$Z$};
        \draw (\ri+.9*\l,-\l/2-\di,-\l/2) node[fill=none] {\footnotesize$Z$};
        \draw (\ri+.45*\l,-\l/2+.45*\l-\di,-\l/2) node[fill=none] {\footnotesize$Z$};
        \draw (\ri+.45*\l,-\l/2-.45*\l-\di,-\l/2) node[fill=none] {\footnotesize$Z$};
        
        \draw (\riii,-\l/2-\di,-\l/2) node[fill=none] {\footnotesize$Z$};
        \draw (\riii+.9*\l,-\l/2-\di,-\l/2) node[fill=none] {\footnotesize$Z$};
        \draw (\riii+.45*\l,-\l/2-\di,-\l/2+.55*\l) node[fill=none] {\footnotesize$Z$};
        \draw (\riii+.45*\l,-\l/2-\di,-\l/2-.55*\l) node[fill=none] {\footnotesize$Z$};
        
        \draw (\rii+.2*\l,-\l/2+.45*\l-\di,-\l/2) node[fill=none] {\footnotesize$Z$};
        \draw (\rii+.2*\l,-\l/2-.45*\l-\di,-\l/2) node[fill=none] {\footnotesize$Z$};
        \draw (\rii+.2*\l,-\l/2-\di,-\l/2+.55*\l) node[fill=none] {\footnotesize$Z$};
        \draw (\rii+.2*\l,-\l/2-\di,-\l/2-.55*\l) node[fill=none] {\footnotesize$Z$};

    \end{tikzpicture}
    \begin{minipage}{.33\columnwidth}
    {\bf (a)}
    \end{minipage}
    \begin{minipage}{.65\columnwidth}
    {\bf (b)}
    \end{minipage}
    \caption{{\bf (a)} Cube and {\bf (b)} cross operators of the X-cube model Hamiltonian on a cubic lattice.}
    \label{Xc-T3-H}
\end{figure}

The X-cube model, as first discussed in \refcite{Sagar16}, is defined on a cubic lattice with qubit degrees of freedom on the edges.
The Hamiltonian
\begin{equation}
H = -\sum_v \left(A_v^{x}+A_v^{y}+A_v^{z}\right) -\sum_c B_c   
\label{eq:H}
\end{equation}
contains two types of terms: cube terms $B_c$ which are products of the twelve Pauli $X$ operators around a cube $c$, and cross terms $A^\mu_v$ which are products of the four Pauli $Z$ operators at a vertex $v$ in the plane normal to the $\mu$-direction where $\mu=x,y,\text{ or }z$ (\figref{Xc-T3-H}).
These terms mutually commute and their energies can be minimized simultaneously.
Moreover, they can be viewed as stabilizer generators for a quantum error-correcting stabilizer code \cite{Gottesman97} whose code space coincides with the Hamiltonian ground space.
One particular ground state is given by $\ket{\psi}=\prod_c (1+B_c)\ket{0}$, where $\ket{0}$ refers to the tensor product of the qubit state $\ket{0}$ on each edge.

Consider an $L_x\times L_y\times L_z$ cubic lattice with periodic boundary conditions.
While there are $3L_xL_yL_z$ qubits in the system and $4L_xL_yL_z$ local terms in the Hamiltonian, the ground state is far from unique.
In fact, the ground state degeneracy (GSD) scales linearly with the size of the system in all three directions:
\begin{equation}
    \log_2{\textrm{GSD}}=2L_x+2L_y+2L_z-3.
    \label{eq:GSDtorus}
\end{equation}


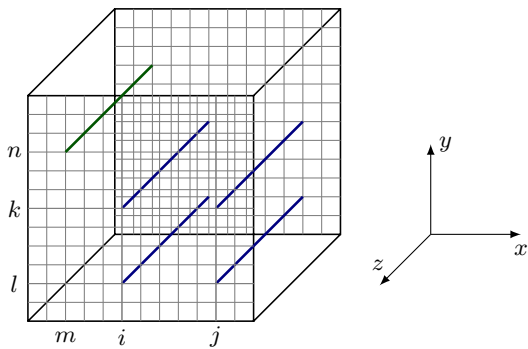
\begin{figure}
    \centering
    \begin{tikzpicture}
        \pgfmathsetmacro{\l}{3}
        \draw[line width=.6]
            (-\l,-\l,0) --++ (0,0,-\l)
            (-\l,0,0) --++ (0,0,-\l)
            (0,-\l,0) --++ (0,0,-\l)
            (0,0,0) --++ (0,0,-\l)
            (0,0,0) -- ++(-\l,0,0) -- ++(0,-\l,0) -- ++(\l,0,0) -- cycle
            (0,0,-\l) -- ++(-\l,0,0) -- ++(0,-\l,0) -- ++(\l,0,0) -- cycle;
        \draw[line width=.4,color=gray]
            (-1*\l/12,-\l,-\l)--++(0,\l,0)
            (-2*\l/12,-\l,-\l)--++(0,\l,0)
            (-3*\l/12,-\l,-\l)--++(0,\l,0)
            (-4*\l/12,-\l,-\l)--++(0,\l,0)
            (-5*\l/12,-\l,-\l)--++(0,\l,0)
            (-6*\l/12,-\l,-\l)--++(0,\l,0)
            (-7*\l/12,-\l,-\l)--++(0,\l,0)
            (-8*\l/12,-\l,-\l)--++(0,\l,0)
            (-9*\l/12,-\l,-\l)--++(0,\l,0)
            (-10*\l/12,-\l,-\l)--++(0,\l,0)
            (-11*\l/12,-\l,-\l)--++(0,\l,0)
        ;
        \draw[line width=.4,color=gray]
            (-\l,-1*\l/12,-\l)--++(\l,0,0)
            (-\l,-2*\l/12,-\l)--++(\l,0,0)
            (-\l,-3*\l/12,-\l)--++(\l,0,0)
            (-\l,-4*\l/12,-\l)--++(\l,0,0)
            (-\l,-5*\l/12,-\l)--++(\l,0,0)
            (-\l,-6*\l/12,-\l)--++(\l,0,0)
            (-\l,-7*\l/12,-\l)--++(\l,0,0)
            (-\l,-8*\l/12,-\l)--++(\l,0,0)
            (-\l,-9*\l/12,-\l)--++(\l,0,0)
            (-\l,-10*\l/12,-\l)--++(\l,0,0)
            (-\l,-11*\l/12,-\l)--++(\l,0,0)
        ;
        \draw[line width=1,color={rgb:blue,1;black,1}]
            (-7*\l/12,-6*\l/12,0)--++(0,0,-\l)
            (-2*\l/12,-6*\l/12,0)--++(0,0,-\l)
            (-7*\l/12,-10*\l/12,0)--++(0,0,-\l)
            (-2*\l/12,-10*\l/12,0)--++(0,0,-\l)
        ;
        \draw[line width=.4,color=gray]
            (-7*\l/12,0,0)--++(0,-\l,0)
            (-2*\l/12,0,0)--++(0,-\l,0)
            (0,-6*\l/12,0)--++(-\l,0,0)
            (0,-10*\l/12,0)--++(-\l,0,0)
        ;
        \draw[line width=.4,color=gray]
            (-1*\l/12,-\l,0)--++(0,\l,0)
            (-2*\l/12,-\l,0)--++(0,\l,0)
            (-3*\l/12,-\l,0)--++(0,\l,0)
            (-4*\l/12,-\l,0)--++(0,\l,0)
            (-5*\l/12,-\l,0)--++(0,\l,0)
            (-6*\l/12,-\l,0)--++(0,\l,0)
            (-7*\l/12,-\l,0)--++(0,\l,0)
            (-8*\l/12,-\l,0)--++(0,\l,0)
            (-9*\l/12,-\l,0)--++(0,\l,0)
            (-10*\l/12,-\l,0)--++(0,\l,0)
            (-11*\l/12,-\l,0)--++(0,\l,0)
        ;
        \draw[line width=.4,color=gray]
            (-\l,-1*\l/12,0)--++(\l,0,0)
            (-\l,-2*\l/12,0)--++(\l,0,0)
            (-\l,-3*\l/12,0)--++(\l,0,0)
            (-\l,-4*\l/12,0)--++(\l,0,0)
            (-\l,-5*\l/12,0)--++(\l,0,0)
            (-\l,-6*\l/12,0)--++(\l,0,0)
            (-\l,-7*\l/12,0)--++(\l,0,0)
            (-\l,-8*\l/12,0)--++(\l,0,0)
            (-\l,-9*\l/12,0)--++(\l,0,0)
            (-\l,-10*\l/12,0)--++(\l,0,0)
            (-\l,-11*\l/12,0)--++(\l,0,0)
        ;
        \draw[line width=1,color={rgb:green,.5;black,1}]
            (-10*\l/12,-3*\l/12,0)--++(0,0,-\l)
        ;
        
        \draw[arrows={-latex}] (\l*.4,-\l,-\l) -- ++(\l*.4,0,0);
        \draw[arrows={-latex}] (\l*.4,-\l,-\l) -- ++(0,\l*.4,0);
        \draw[arrows={-latex}] (\l*.4,-\l,-\l) -- ++(0,0,\l/1.7);
        
        \draw (\l*.4+.2,-\l+1.2,-\l) node[fill=none] {$y$};
        \draw (\l-.6,-\l-.2,-\l) node[fill=none] {$x$};
        \draw (\l*.4-.7,-\l-.4,-\l) node[fill=none] {$z$};
        
        \draw(-\l/12*7,-\l*1.07,0) node[fill=none] {$i$};
        \draw(-\l/12*2,-\l*1.07,0) node[fill=none] {$j$};
        \draw(-\l/12*10,-\l*1.07,0) node[fill=none] {$m$};
        
        \draw(-\l*1.06,-\l/12*3,0) node[fill=none] {$n$};
        \draw(-\l*1.06,-\l/12*6,0) node[fill=none] {$k$};
        \draw(-\l*1.06,-\l/12*10,0) node[fill=none] {$l$};
    \end{tikzpicture}
    \caption{Visualization of logical operators. The green string corresponds to $W^z_{mn}$. The product of the four operators corresponding to the blue strings is equal to the identity, as described in the main text.}
    \label{fig:logicalOperators}
\end{figure}

There are hence a large number of `logical operators' that commute with all of the terms in the Hamiltonian and map one ground state to another.\cite{Slagle17,BernevigEntropy}
An over-complete set of $X$-type logical operators is given by the set of closed string-like operators $W^\mu_{ij}$,
which is a product of $X$ operators over all $\mu$-oriented edges with coordinates $(i,j)$ in the plane normal to $\mu$ (see \figref{fig:logicalOperators}).
This set is over-complete in the sense that products of the form $W^\mu_{ij}W^\mu_{il}W^\mu_{kl}W^\mu_{kj}$ are equal to a product of some $B_c$ cube operators, and thus act trivially on the ground state manifold (here the four sets of coordinates lie at the corners of a rectangle in the plane normal to $\mu$, as shown in \figref{fig:logicalOperators}).
There are $L_xL_y+L_yL_z+L_zL_x-2L_x-2L_y-2L_z+3$ such relations corresponding to unique products of cube operators, thus implying \eqnref{eq:GSDtorus}. Moreover, it was found that for each ground state, the entanglement entropy of a region $R$ satisfies an area law with subleading corrections linear in the length of $R$, which has a similar origin as the subextensive scaling of ground state degeneracy.\cite{HermeleEntropy,BernevigEntropy}

\begin{figure}
    \centering
    \begin{tikzpicture}
        \pgfmathsetmacro{\l}{.7}
        \draw[line width=.4, color=gray]
            (0,0,0) --++ (0,0,-\l)
            (0,-1*\l,0) --++ (0,0,-\l)
            (0,-2*\l,0) --++ (0,0,-\l)
            (0,-3*\l,0) --++ (0,0,-\l)
            (0,-4*\l,0) --++ (0,0,-\l)
            (0,-5*\l,0) --++ (0,0,-\l)
            (-1*\l,0,0) --++ (0,0,-\l)
            (-1*\l,-1*\l,0) --++ (0,0,-\l)
            (-1*\l,-2*\l,0) --++ (0,0,-\l)
            (-1*\l,-3*\l,0) --++ (0,0,-\l)
            (-1*\l,-4*\l,0) --++ (0,0,-\l)
            (-1*\l,-5*\l,0) --++ (0,0,-\l)
            (-2*\l,0,0) --++ (0,0,-\l)
            (-2*\l,-1*\l,0) --++ (0,0,-\l)
            (-2*\l,-2*\l,0) --++ (0,0,-\l)
            (-2*\l,-3*\l,0) --++ (0,0,-\l)
            (-2*\l,-4*\l,0) --++ (0,0,-\l)
            (-2*\l,-5*\l,0) --++ (0,0,-\l)
            (-3*\l,0,0) --++ (0,0,-\l)
            (-3*\l,-1*\l,0) --++ (0,0,-\l)
            (-3*\l,-2*\l,0) --++ (0,0,-\l)
            (-3*\l,-3*\l,0) --++ (0,0,-\l)
            (-3*\l,-4*\l,0) --++ (0,0,-\l)
            (-3*\l,-5*\l,0) --++ (0,0,-\l)
            (-4*\l,0,0) --++ (0,0,-\l)
            (-4*\l,-1*\l,0) --++ (0,0,-\l)
            (-4*\l,-2*\l,0) --++ (0,0,-\l)
            (-4*\l,-3*\l,0) --++ (0,0,-\l)
            (-4*\l,-4*\l,0) --++ (0,0,-\l)
            (-4*\l,-5*\l,0) --++ (0,0,-\l)
            (-5*\l,0,0) --++ (0,0,-\l)
            (-5*\l,-1*\l,0) --++ (0,0,-\l)
            (-5*\l,-2*\l,0) --++ (0,0,-\l)
            (-5*\l,-3*\l,0) --++ (0,0,-\l)
            (-5*\l,-4*\l,0) --++ (0,0,-\l)
            (-5*\l,-5*\l,0) --++ (0,0,-\l)
            (-6*\l,0,0) --++ (0,0,-\l)
            (-6*\l,-1*\l,0) --++ (0,0,-\l)
            (-6*\l,-2*\l,0) --++ (0,0,-\l)
            (-6*\l,-3*\l,0) --++ (0,0,-\l)
            (-6*\l,-4*\l,0) --++ (0,0,-\l)
            (-6*\l,-5*\l,0) --++ (0,0,-\l)
            (-7*\l,0,0) --++ (0,0,-\l)
            (-7*\l,-1*\l,0) --++ (0,0,-\l)
            (-7*\l,-2*\l,0) --++ (0,0,-\l)
            (-7*\l,-3*\l,0) --++ (0,0,-\l)
            (-7*\l,-4*\l,0) --++ (0,0,-\l)
            (-7*\l,-5*\l,0) --++ (0,0,-\l)
            
            (0,0,0)--++(-\l*8,0,0)
            (0,-1*\l,0)--++(-\l*8,0,0)
            (0,-2*\l,0)--++(-\l*8,0,0)
            (0,-3*\l,0)--++(-\l*8,0,0)
            (0,-4*\l,0)--++(-\l*8,0,0)
            (-1*\l,\l,0)--++(0,-\l*6,0)
            (-2*\l,\l,0)--++(0,-\l*6,0)
            (-3*\l,\l,0)--++(0,-\l*6,0)
            (-4*\l,\l,0)--++(0,-\l*6,0)
            (-5*\l,\l,0)--++(0,-\l*6,0)
            (-6*\l,\l,0)--++(0,-\l*6,0)
            (-7*\l,\l,0)--++(0,-\l*6,0)
            (0,0,-\l)--++(-\l*7,0,0)
            (0,-1*\l,-\l)--++(-\l*8,0,0)
            (0,-2*\l,-\l)--++(-\l*8,0,0)
            (0,-3*\l,-\l)--++(-\l*8,0,0)
            (0,-4*\l,-\l)--++(-\l*8,0,0)
            (-1*\l,\l,-\l)--++(0,-\l*6,0)
            (-2*\l,\l,-\l)--++(0,-\l*6,0)
            (-3*\l,\l,-\l)--++(0,-\l*6,0)
            (-4*\l,\l,-\l)--++(0,-\l*6,0)
            (-5*\l,\l,-\l)--++(0,-\l*6,0)
            (-6*\l,\l,-\l)--++(0,-\l*6,0)
            (-7*\l,\l,-\l)--++(0,-\l*6,0)
            
            (-0*\l,\l,0)--++(0,0,-\l)
            (-1*\l,\l,0)--++(0,0,-\l)
            (-2*\l,\l,0)--++(0,0,-\l)
            (-3*\l,\l,0)--++(0,0,-\l)
            (-4*\l,\l,0)--++(0,0,-\l)
            (-5*\l,\l,0)--++(0,0,-\l)
            (-6*\l,\l,0)--++(0,0,-\l)
            (-7*\l,\l,0)--++(0,0,-\l)
            (-8*\l,\l,0)--++(0,0,-\l)
            (-8*\l,0*\l,0)--++(0,0,-\l)
            (-8*\l,-1*\l,0)--++(0,0,-\l)
            (-8*\l,-2*\l,0)--++(0,0,-\l)
            (-8*\l,-3*\l,0)--++(0,0,-\l)
            (-8*\l,-4*\l,0)--++(0,0,-\l)
            (-8*\l,-5*\l,0)--++(0,0,-\l)
            
            (0,\l,0)--++(-8*\l,0,0)--++(0,-6*\l,0)--++(8*\l,0,0)--cycle
            (0,\l,-\l)--++(-8*\l,0,0)--++(0,-6*\l,0)--++(8*\l,0,0)--cycle
            ;
            
            \draw[line width=.5]
            (-2*\l,+\l,0)--++(-\l,0,0)--++(0,-\l,0)--++(\l,0,0)--cycle
            (-7*\l,+\l,0)--++(-\l,0,0)--++(0,-\l,0)--++(\l,0,0)--cycle
            (-2*\l,+\l,-\l)--++(-\l,0,0)--++(0,-\l,0)--++(\l,0,0)--cycle
            (-7*\l,+\l,-\l)--++(-\l,0,0)--++(0,-\l,0)--++(\l,0,0)--cycle
            (-2*\l,-3*\l,0)--++(-\l,0,0)--++(0,-\l,0)--++(\l,0,0)--cycle
            (-7*\l,-3*\l,0)--++(-\l,0,0)--++(0,-\l,0)--++(\l,0,0)--cycle
            (-2*\l,-3*\l,-\l)--++(-\l,0,0)--++(0,-\l,0)--++(\l,0,0)--cycle
            (-7*\l,-3*\l,-\l)--++(-\l,0,0)--++(0,-\l,0)--++(\l,0,0)--cycle
            (-8*\l,\l,0)--++(0,0,-\l)
            (-8*\l,0,0)--++(0,0,-\l)
            (-7*\l,\l,0)--++(0,0,-\l)
            (-2*\l,\l,0)--++(0,0,-\l)
            (-2*\l,0,0)--++(0,0,-\l)
            (-3*\l,\l,0)--++(0,0,-\l)
            (-8*\l,-3*\l,0)--++(0,0,-\l)
            (-8*\l,-4*\l,0)--++(0,0,-\l)
            (-7*\l,-4*\l,0)--++(0,0,-\l)
            (-2*\l,-3*\l,0)--++(0,0,-\l)
            (-2*\l,-4*\l,0)--++(0,0,-\l)
            (-3*\l,-4*\l,0)--++(0,0,-\l)
            ;
            
            \draw[line width=1.2,color={rgb:red,1;black,.75}]
            (-\l*6,-\l*1,0)--++(0,0,-\l)
            (-\l*5,-\l*1,0)--++(0,0,-\l)
            (-\l*4,-\l*1,0)--++(0,0,-\l)
            (-\l*3,-\l*1,0)--++(0,0,-\l)
            (-\l*7,-\l*1,0)--++(0,0,-\l)
            (-\l*6,-\l*2,0)--++(0,0,-\l)
            (-\l*5,-\l*2,0)--++(0,0,-\l)
            (-\l*4,-\l*2,0)--++(0,0,-\l)
            (-\l*3,-\l*2,0)--++(0,0,-\l)
            (-\l*7,-\l*2,0)--++(0,0,-\l)
            (-\l*6,-\l*3,0)--++(0,0,-\l)
            (-\l*5,-\l*3,0)--++(0,0,-\l)
            (-\l*4,-\l*3,0)--++(0,0,-\l)
            (-\l*3,-\l*3,0)--++(0,0,-\l)
            (-\l*7,-\l*3,0)--++(0,0,-\l)
            (-\l*6,0,0)--++(0,0,-\l)
            (-\l*5,0,0)--++(0,0,-\l)
            (-\l*4,0,0)--++(0,0,-\l)
            (-\l*3,0,0)--++(0,0,-\l)
            (-\l*7,0,0)--++(0,0,-\l)
            ;
            
            \draw[line width=1,color={rgb:blue,1;black,1}]
            (-\l*1,-\l*5,0)--++(0,\l*5,0)
            ;
            
            \draw[black,fill=black] (-\l,0,0) circle (.03);
            \draw[black,fill=black] (-\l,-\l*5,0) circle (.03);
            
    \end{tikzpicture}
    \caption{Visualization of particle creation operators. The red links correspond to a membrane geometry on the dual lattice. The product of $Z$ operators over these edges excites the (darkened) cube operators at the corners. The product of $X$ operators over the links comprising the straight open blue string creates excitations at its endpoints (black dots).}
    \label{fig:excitations}
\end{figure}
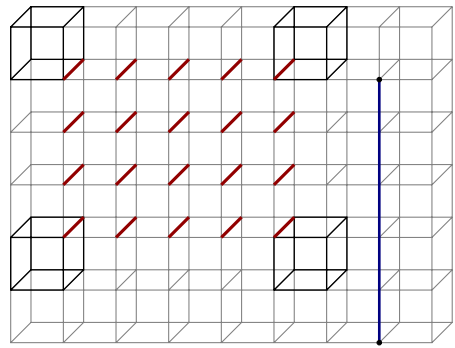

Logical operators correspond to processes where particle anti-particle pairs are created out of the vacuum, wound around the torus, and then annihilated. Straight open string operators $W^\mu_{ij}\left(\mu_1,\mu_2\right)$ anti-commute with the vertex Hamiltonian terms at the endpoints $\mu_1$ and $\mu_2$, corresponding to excitations which live on the vertices of the lattice. Here $W^\mu_{ij}\left(\mu_1,\mu_2\right)$ is defined to be the product of $X$ operators over $\mu$-oriented edges between $\mu=\mu_1$ and $\mu=\mu_2$ with coordinate $(i,j)$ in the plane normal to $\mu$ (see \figref{fig:excitations}). Conversely, acting with bent string operators introduces additional energetic costs at the corners. Therefore the particles living at the endpoints of straight open strings are energetically confined to live on a line; in this sense, they are dimension-1 particles. \cite{Sagar16} These particles obey an unconventional fusion rule: triples of particles living along $x$-, $y$-, and $z$-oriented lines may annihilate into the vacuum. On the other hand, acting with a closed string 
operator around a rectangle creates an excitation at each corner of the rectangle. A pair of particles at adjacent corners may be viewed as a single dipole-like object which is itself a dimension-2 particle and is mobile in the plane normal to the edges connecting the two corners.

In addition to these string-like operators, there are membrane-like operators which are products of $Z$ operators over qubits corresponding to a membrane geometry on the dual lattice (see \figref{fig:excitations}). A rectangular membrane operator anti-commutes with the cube Hamiltonian terms at its corners. A pair of adjacent corner excitations created by a rectangular membrane operator is likewise a dimension-2 dipolar particle, free to move in a plane perpendicular to its moment. A process whereby a pair of such membrane dipoles is created, separated, wound around the torus and annihilated, corresponds to a string-like $Z$-type logical operator.

\section{X-cube Model on Generic Lattices}
\label{sec:lattices}

The 3D toric code model, which represents the traditional 3D $\mathbb{Z}_2$ gauge theory topological order, can be defined on any lattice on any manifold. For the $X$-cube model, however, it is not clear if this is possible. In this section, we explain how to define the $X$-cube model on a special class of lattices, which will enable constructions on general spatial 3-manifolds in the subsequent section. 

\subsection{Intersecting leaves}
\label{sec:leaves}

We construct a lattice by embedding a large number of transversely intersecting surfaces, referred to as leaves, into the 3-manifold $M$. Vertices of the resulting cellulation lie at triple intersection points of leaves, while edges lie along the intersections of pairs of leaves; a qubit is placed on each edge. We assume that the location of the leaves are generic enough such that no three leaves intersect along the same line.
The cubic lattice on the 3-torus can be viewed in this way as three orthogonal stacks of toroidal leaves---the $xy$, $yz$, and $xz$ planes of $T^3={\mathbb{R}^3}/{\mathbb{Z}^3}$.
Unlike the cubic lattice, the general construction may result in some number of non-cubical 3-cells.
Crucially, however, every vertex in this type of cellulation is locally isomorphic to a cubic lattice vertex.
This fact allows the X-cube Hamiltonian to be defined as per \eqnref{eq:H}. 
Similar to the cubic lattice, the three cross operators $A^\mu_v$ are products of $Z$ operators over the four edges emanating from $v$ in the leaf labeled by $\mu$.
The $B_c$ operator is in general a product of $X$ operators over all edges of the 3-cell $c$.
The cellulation geometry ensures that the terms in the Hamiltonian are mutually commuting.

The structure of the excitation types and fusion properties carries over from the cubic lattice version of the X-cube model. However, the notion of dimension-1 and dimension-2 particles is revised in a natural way. In the general lattice construction, dimension-1 particles created at the ends of open string operators are freely mobile along the intersection lines of pairs of cellulating surfaces. Furthermore, dimension-2 particles, such as fracton dipoles, are free to move along leaves that are orthogonal to the direction of the dipole moment.
In the general setting, logical operators correspond to processes where particle pairs are created, wound around the intersection circle of two surfaces, and then annihilated. Unlike the three-torus, for general manifolds these circles may be contractible.

\subsection{Spherical leaf construction and 3+1D \texorpdfstring{$\mathbb{Z}_2$}{Z\_2} gauge theory}
\label{sec:spheres}

Before we turn to the notion of total foliation,
we first note the existence of a somewhat anomalous lattice construction that defines an X-cube model on any 3-manifold $M$.
We choose the leaves of the cellulation to be contractible small spheres placed randomly or periodically throughout $M$.
As explained above, a lattice is formed by placing vertices at the intersection of three spheres and edges along the intersection of two spheres.
As long as the spheres are packed closely enough that each sphere intersects with several other spheres,
this construction defines a lattice in $M$.
The X-cube Hamiltonian on this lattice generalizes \eqnref{eq:H},
with three cross terms at each vertex $v$ and a $B_c$ term for each 3-cell.
The operator $B_c$ is a product of $X$ operators over the edges of the 3-cell $c$.

This construction allows the X-cube model to be defined on any manifold. However, the resulting model can have fully mobile deconfined point excitations, and a constant GSD of 8 on the 3-torus. Thus we conclude that the model can exhibit conventional 3+1D $\mathbb{Z}_2$ gauge theory topological order rather than fracton order.
We have numerically verified a GSD of 8 for a 3-torus covered with spherical leaves of radius 0.46 centered at points of an FCC lattice (see \sfigref{fig:logicalSpheres}{a-b}). We used the method equivalent to the one described in Appendix B of \refcite{MaLayers}. The unit cell in this configuration contains 48 links. By enumerating the 48 links in a unit cell and inputting the complicated stabilizer Hamiltonian into the algorithm, we identify a ground state degeneracy of 8. 

The 8-fold GSD can be understood by considering the string and membrane logical operators acting on the topologically protected ground space. A membrane and string operator is shown in \sfigref{fig:logicalSpheres}{d-e}. The membrane operator is a product of $X$ operators on the blue edges, whereas the string operator is a product of $Z$ operators on the red edges. These operators commute with the Hamiltonian, but anticommute with each other, and therefore describe one qubit in the degenerate ground state Hilbert space. 90 degree rotation gives the two other pairs of these operators. 

The deconfined point-like charges of the model correspond to 3-cell excitations lying at the ends of open string operators (as in \sfigref{fig:logicalSpheres}{b}).
These particles are fully mobile because the corresponding string operators can bend without creating additional excitations. This is a surprising result, as excitations of the 3-cell operators on a cubic lattice are immobile fractons. Conversely, violations of the cross operators lie along the boundary of open membrane operators (shown in blue in \sfigref{fig:logicalSpheres}{d-e}).
These excitations correspond to flux loops of the 3+1D $\mathbb{Z}_2$ gauge theory.
Hence, we see that both the fractional excitations and logical operators match those of 3+1D $\mathbb{Z}_2$ gauge theory. Other arrangements of spheres may also result in the 3+1D $\mathbb{Z}_2$ gauge theory. 

\begin{figure}
    \begin{minipage}{.4\columnwidth}
    \includegraphics[width=\columnwidth]{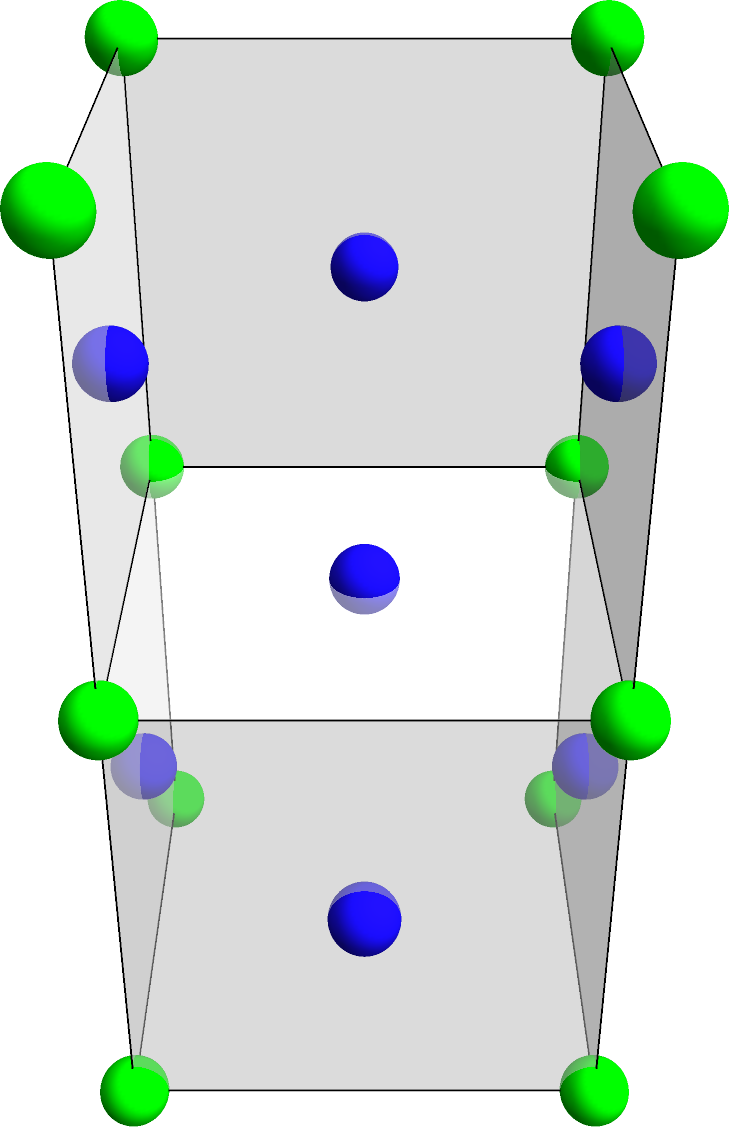}
    \end{minipage}
    \begin{minipage}{.54\columnwidth}
    \includegraphics[width=\columnwidth]{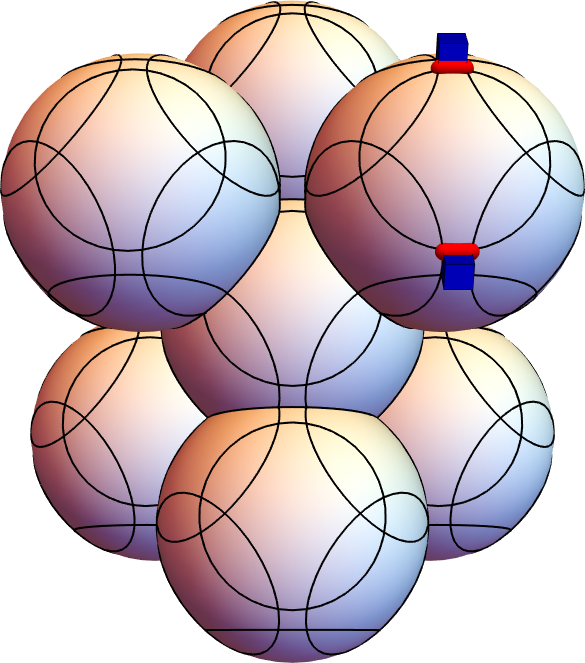}
    \end{minipage} \\ \vspace{.1cm}
    \begin{minipage}{.4\columnwidth}
    {\bf (a)}
    \end{minipage}
    \begin{minipage}{.54\columnwidth}
    {\bf (b)}
    \end{minipage} \\ \vspace{.1cm}
    \includegraphics[width=.22\columnwidth]{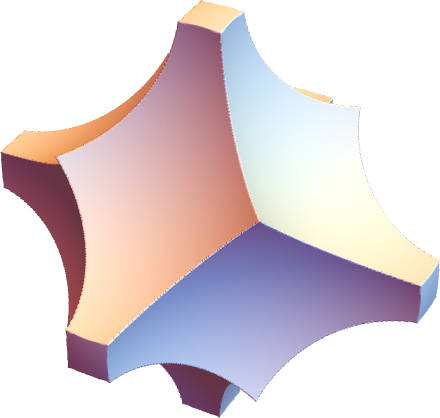}
    \includegraphics[width=.21\columnwidth]{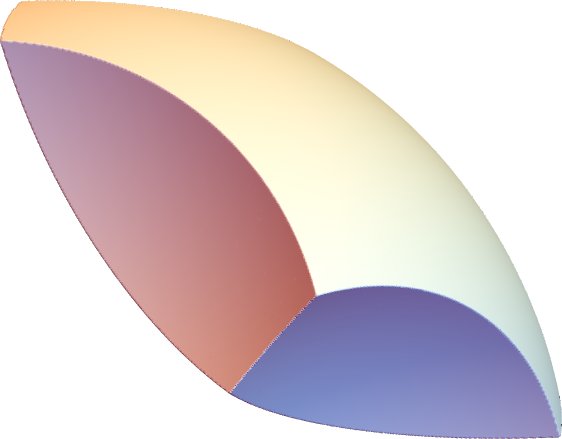}
    \includegraphics[width=.14\columnwidth]{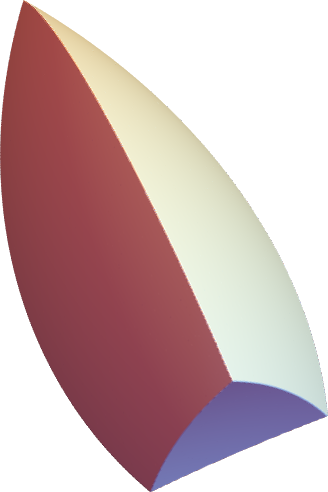}
    \includegraphics[width=.19\columnwidth]{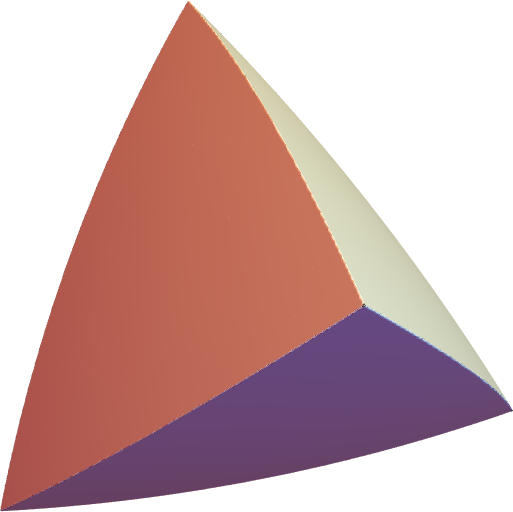}
    \includegraphics[width=.19\columnwidth]{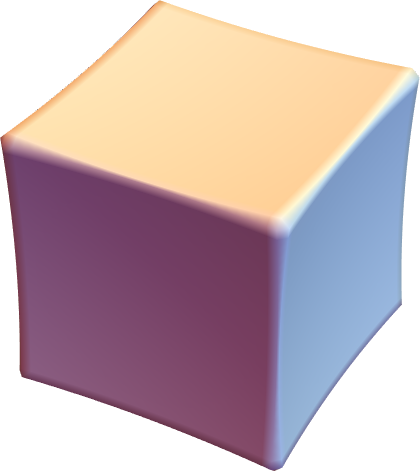} \\
    {\bf (c)} \\ \vspace{.1cm}
    \begin{minipage}{.47\columnwidth}
    \includegraphics[width=\columnwidth]{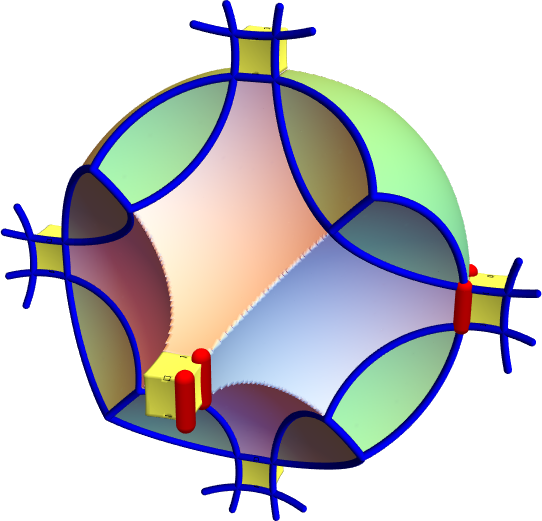}
    \end{minipage} \hspace{.03\columnwidth}
    \begin{minipage}{.47\columnwidth}
    \includegraphics[width=\columnwidth]{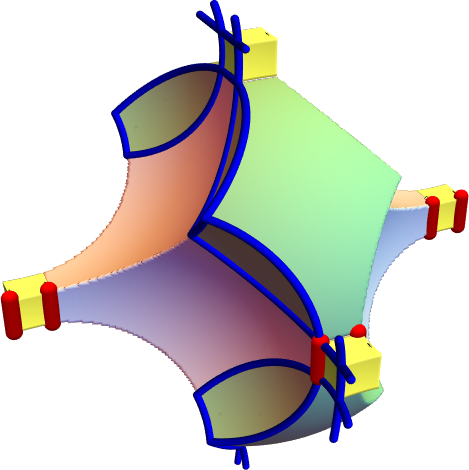}
    \end{minipage} \\ \vspace{.1cm}
    \begin{minipage}{.47\columnwidth}
    {\bf (d)}
    \end{minipage} \hspace{.03\columnwidth}
    \begin{minipage}{.47\columnwidth}
    {\bf (e)}
    \end{minipage}
    \caption{A construction with periodically placed spheres. (\secref{sec:spheres}). {\bf (a-b)} We place spheres of radius 0.46 on an face-centered cubic (FCC) lattice. The spheres in (b) are located at the blue points of the FCC lattice in (a). When the X-cube model is defined on the resulting lattice, the phase is equivalent to the 3D toric code. {\bf (b)} The toric code charges reside on small cubes. These charges can hop e.g. between the two blue cubes via a string of $Z$ operators on the two red edges. {\bf (c)} The elementary 3-cells of the cellulation. {\bf (d-e)} Membrane and string operators. The membrane operator is a product of $X$ operators on the blue edges, whereas the string operator is a product of $Z$ operators on the red edges.}
    \label{fig:logicalSpheres}
\end{figure}

Thus, a different approach must be considered to construct lattices whose X-cube constructions realize fracton order. Since dimension-1 and dimension-2 particles are constrained to move within individual leaves, extended dimension-1 and dimension-2 particles and a robust ground state degeneracy that scales subextensively with system size can be realized only in the presence of leaves that are non-locally embedded in the 3-manifold. This consideration motivates the following section.

\section{X-cube model on general manifolds via total foliation}
\label{sec:manifolds}

\begin{figure}[htbp!]
    \includegraphics[scale=.18]{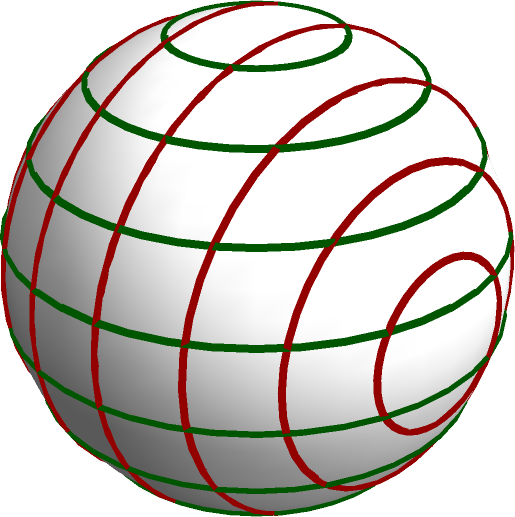}
    \hspace{.2cm}
    \begin{tikzpicture}
        \pgfmathsetmacro{\l}{2.45}
        
        \draw[dashed, line width=.8, color={rgb:black,.75;red,1}] (-\l/2,0,0)--++(0,0,-\l)--++(0,-\l,0)--++(0,0,\l)--cycle;
        \draw[line width=.8, color={rgb:black,.75;red,1}] (-2*\l/3,0,0)--++(0,0,-\l)--++(0,-\l,0)--++(0,0,\l)--cycle;
        \draw[line width=.8, color={rgb:black,.75;red,1}] (-\l/3,0,0)--++(0,0,-\l)--++(0,-\l,0)--++(0,0,\l)--cycle;
        
        \draw[dashed, line width=.8, color={rgb:black,1;green,.5}] (0,-\l/2,0)--++(-\l,0,0)--++(0,0,-\l)--++(\l,0,0)--cycle;
        \draw[line width=.8, color={rgb:black,1;green,.5}] (0,-\l/4,0)--++(-\l,0,0)--++(0,0,-\l)--++(\l,0,0)--cycle;
        \draw[line width=.8, color={rgb:black,1;green,.5}] (0,-3*\l/4,0)--++(-\l,0,0)--++(0,0,-\l)--++(\l,0,0)--cycle;
        
        \draw[line width=.8, color={rgb:black,1;blue,1}] (0,0,-\l/2)--++(-\l,0,0)--++(0,-\l,0)--++(\l,0,0)--cycle;
        
        \draw (-\l/2,-\l/2,0) node[fill=none, color=black] {\Large {\bf Half-twist}};
        \draw (-\l/2,-\l/2,-\l) node[fill=none, color=black] {\Large \begin{turn}{180}{\bf Half-twist}\end{turn}};
        
        \draw[line width=.6]
        (-\l,-\l,0) --++ (0,0,-\l)
        (-\l,0,0) --++ (0,0,-\l)
        (0,-\l,0) --++ (0,0,-\l)
        (0,0,0) --++ (0,0,-\l)
        (0,0,0) -- ++(-\l,0,0) -- ++(0,-\l,0) -- ++(\l,0,0) -- cycle
        (0,0,-\l) -- ++(-\l,0,0) -- ++(0,-\l,0) -- ++(\l,0,0) -- cycle;
        
    \end{tikzpicture} \\
    \begin{minipage}{.4\columnwidth}
    {\bf (a)}
    \end{minipage}
    \begin{minipage}{.49\columnwidth}
    {\bf (c)}
    \end{minipage} \\\vspace{.4cm}
    
    \includegraphics[scale=.18]{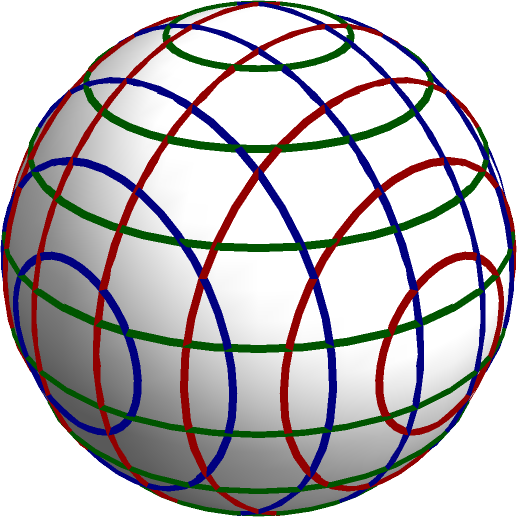}
    \hspace{.2cm}
    \begin{tikzpicture}
        \pgfmathsetmacro{\l}{2.45}
        
        \draw[dashed, line width=.8, color={rgb:black,.75;red,1}] (-\l/2,0,0)--++(0,0,-\l)--++(0,-\l,0)--++(0,0,\l)--cycle;
        \draw[line width=.8, color={rgb:black,.75;red,1}] (-2*\l/3,0,0)--++(0,0,-\l)--++(0,-\l,0)--++(0,0,\l)--cycle;
        \draw[line width=.8, color={rgb:black,.75;red,1}] (-\l/3,0,0)--++(0,0,-\l)--++(0,-\l,0)--++(0,0,\l)--cycle;
        
        \draw[line width=.8, color={rgb:black,1;green,.5}] (0,-\l/2,0)--++(-\l,0,0)--++(0,0,-\l)--++(\l,0,0)--cycle;
        
        \draw[line width=.8, color={rgb:black,1;blue,1}] (0,0,-\l/2)--++(-\l,0,0)--++(0,-\l,0)--++(\l,0,0)--cycle;
        
        \draw (-\l/2,-\l/2,0) node[fill=none, color=black] {\LARGE $\boldsymbol{K^2\times S^1}$};
        \draw (-\l/2,-\l/2,-\l) node[fill=none, color=black] {\LARGE \reflectbox{$\boldsymbol{K^2\times S^1}$}};
        
        \draw[line width=.6]
        (-\l,-\l,0) --++ (0,0,-\l)
        (-\l,0,0) --++ (0,0,-\l)
        (0,-\l,0) --++ (0,0,-\l)
        (0,0,0) --++ (0,0,-\l)
        (0,0,0) -- ++(-\l,0,0) -- ++(0,-\l,0) -- ++(\l,0,0) -- cycle
        (0,0,-\l) -- ++(-\l,0,0) -- ++(0,-\l,0) -- ++(\l,0,0) -- cycle;
        
    \end{tikzpicture}
    \begin{minipage}{.4\columnwidth}
    {\bf (b)}
    \end{minipage}
    \begin{minipage}{.49\columnwidth}
    {\bf (d)}
    \end{minipage}
    \\\vspace{.4cm}
    \includegraphics[scale=1]{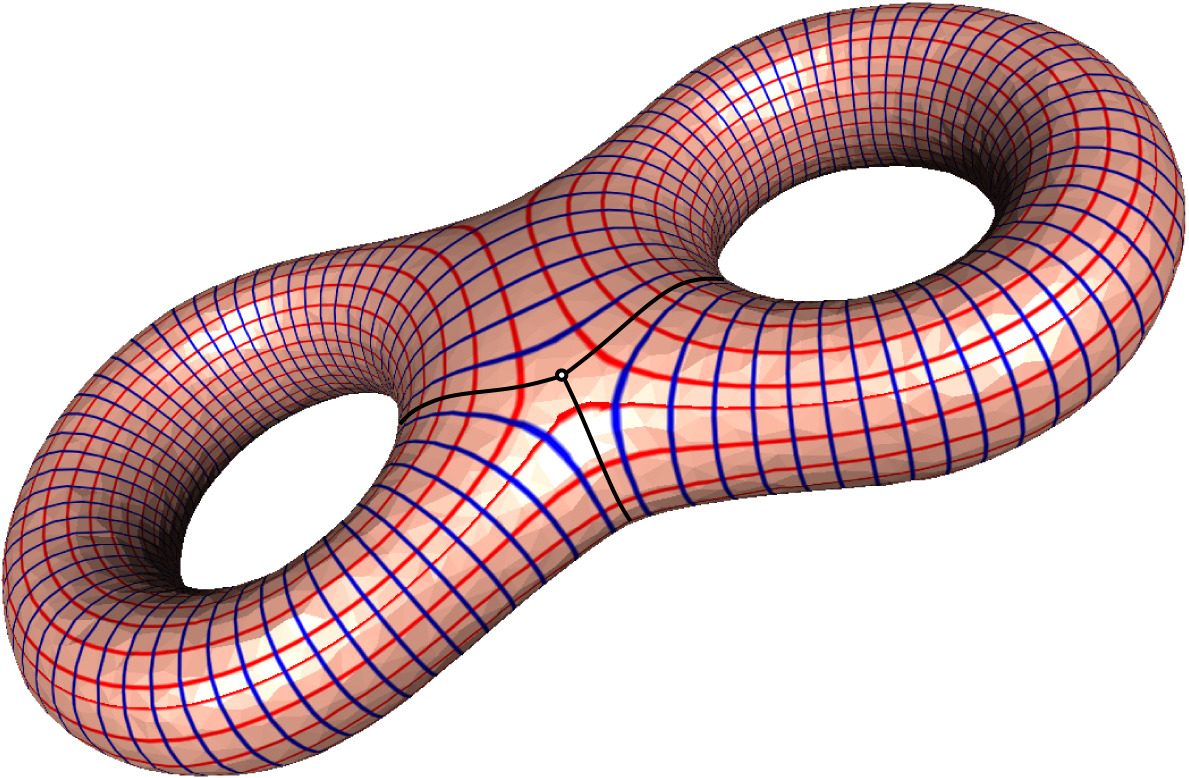}
    \begin{minipage}{.8\columnwidth}
    {\bf (e)}
    \end{minipage}
    \caption{{\bf (a)} A spherical cross-section of a cellulation of $S^2\times S^1$ with $L_x=L_y=8$. {\bf (b)} The $t=0$ equator  of $S^3$ defined as the locus of points in $\mathbb{R}^4$ satisfying $x^2+y^2+z^2+t^2=1$. In this example, $S^3$ is foliated by 8 spherical leaves of constant $x$, $y$, and $z$, which are colored red, green, and blue. Although the sphere drawn in (a) is a leaf, the sphere drawn in (b) is not a leaf; it is merely a convenient cross-section. {\bf (c)} The half-twist manifold, constructed by identifying opposite faces of a cube. The front and back faces are glued after a $180\degree$ twist. The
    dashed red and green squares are outlines of embedded Klein bottles. The pair of solid red (or green) squares outline a single torus, as does the blue square. {\bf (d)} The 3-manifold $K^2\times S^1$, viewed as a cube with opposite faces identified; front and back faces are identified after a reflection across the vertical bisector. The pair of solid red squares outlines a single embedded torus, as do the dashed red square and solid blue square. The solid green square outlines an embedded Klein bottle. {\bf  (e)} Figure courtesy of \refcite{Hexahedral}. A $\Sigma_2$ cross-section of a cellulation of $\Sigma_2\times S^1$. The red and blue lines correspond to leaves of respective singular foliations. The singularities are indicated by the black lines.}
    \label{fig:manifolds}
\end{figure}
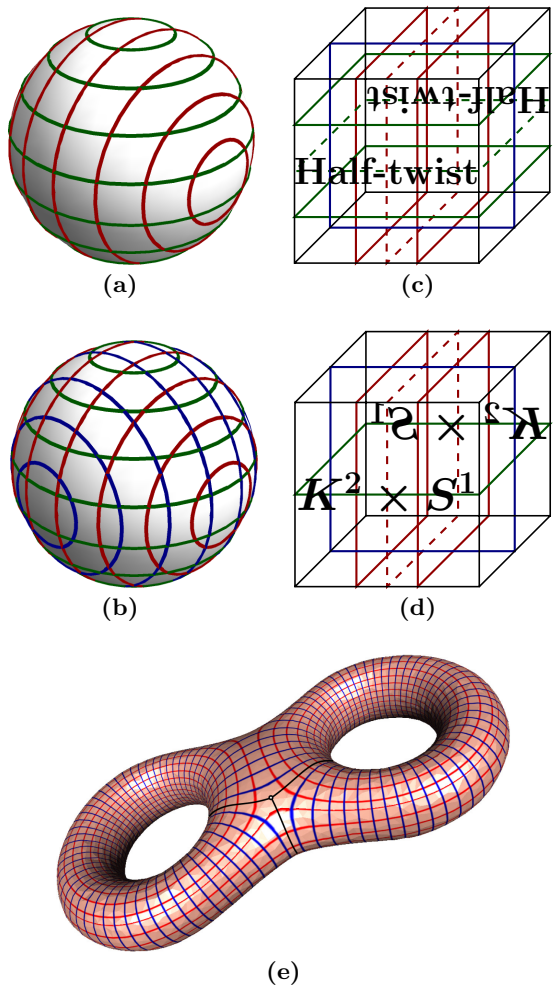

In this section, we consider cellulations of a 3-manifold $M$ constructed by embedding into $M$ three transversely intersecting stacks of parallel surfaces,
which are assigned $x$, $y$, and $z$ labels and are
composed of $L_x$, $L_y$, and $L_z$ layers, respectively.
Each stack of surfaces may be viewed as a discrete sample of compact leaves of a (possibly singular)\footnote{In a singular foliation, leaves at singularities may differ in dimension (i.e. dimension less than two in the case of 3-manifolds).} two-dimensional foliation of $M$.
(A $p$-dimensional foliation of a manifold $M$ is an infinite partition of $M$ into a collection of disjoint parallel $p$-dimensional submanifolds of $M$ with infinitesimal separation. The submanifolds are referred to as leaves.)
This approach is reminiscent of the mathematical notion of total foliation. \footnote{A total foliation of an $n$-dimensional manifold consists of $n$ sets of $(n-1)$-dimensional foliations that are transverse at every point.} However, our construction differs in that we allow the foliations to be singular (containing leaves that are of a different dimension) but require that the leaves are compact so that the resulting lattice has a finite number of edges.
The discrete foliations are required to obey the following transversality conditions, which can always be satisfied by an appropriate choice of leaves: pairs of foliating surfaces must intersect transversely (i.e. not tangentially), and triples of surfaces must intersect at points. 
These generalized cellulations retain a notion of continuum limit, as they can be arbitrarily refined by adding leaves to any of the three constituent foliations.

We find that the ground state degeneracy of the generalized X-cube model obeys the formula
\begin{equation}
    \log_2\textrm{GSD}=b_xL_x+b_yL_y+b_zL_z-c
    \label{eq:GSDgenus}
\end{equation}
where $b_\mu$ is the first Betti number with $\mathbb{Z}_2$ coefficients
\cite{foot:Betti} of the surfaces comprising the $\mu$-oriented foliation\footnote{If the surfaces in the $\mu$-oriented foliation have different Betti numbers, then one should instead multiply by the corresponding Betti number for each surface.}, and $c$ is a constant sensitive to the topology of the intersections of the three foliations.
As we will discuss in \secref{sec:RG}, the presence of the first three terms can be understood in terms of an entanglement renormalization transformation which grows the system size by adding layers of toric code states, which have $\log_2\text{GSD} = b$.

We stress that the degenerate ground space is sensitive to the foliation structure imposed on the 3-manifold as well as its topology, and that it is possible to endow the same 3-manifold with differing foliation structures (for example in the case of the half-twist manifold discussed below). Furthermore, we note that singularities in the foliation structure may result in partial splitting of the GSD (in the presence of local perturbations) due to the existence of logical operators with local support. This occurs, for instance in the case of $S^2\times S^1$, in which two of the foliations have point singularities. Conversely, in the case of $\Sigma_2\times S^1$ depicted in \sfigref{fig:manifolds}{e}, the foliations exhibit codimension-1 singularities, but the resulting models do not contain logical operators with local support.

We now turn to some examples. The results are summarized in Table \ref{tab:GSD}.
We have numerically verified the expressions for GSD using a method equivalent to the one described in Appendix B of \refcite{MaLayers}.

\begin{table*}
\begin{center}
\begin{tabular}{c|c|c|c|l|c}
3-manifold & $x$-leaves & $y$-leaves & $z$-leaves & $\quad\quad\;\log_2 \text{GSD}$ & $c$ \\\hline
$T^3$                   & $L_x \times T^2$ & $L_y \times T^2$ & $L_z \times T^2$
                        & $2L_x + 2L_y +2L_z- 3$        & $3$ \\
$S^2 \times S^1$        & $L_x \times T^2$ & $L_y \times T^2$ & $L_z \times S^2$
                        & $2L_x + 2L_y - 1 \quad\quad\;\, *$ & $1$ \\
$S^3$                   & $L_x \times S^2$ & $L_y \times S^2$ & $L_z \times S^2$
                        & $0$                      & $0$ \\
\text{half-twist}       & $L_x \times T^2$ & $L_y \times T^2$ & $L_z \times T^2$
                        & $2L_x + 2L_y + 2L_z$     & $0$ \\
\text{half-twist}       & $(L_x-1) \times T^2 + K^2$ & $L_y \times T^2$ & $L_z \times T^2$
                        & $2L_x + 2L_y + 2L_z - 2$ & $2$ \\
\text{half-twist}       & $(L_x-1) \times T^2 + K^2$ & $(L_y-1) \times T^2 + K^2$ & $L_z \times T^2$
                        & $2L_x + 2L_y + 2L_z - 3$ & $3$ \\
$K^2 \times S^1$        & $L_x \times T^2$ & $L_y \times T^2$ & $L_z \times K^2$
                        & $2L_x + 2L_y + 2L_z - 2$ & $2$ \\
$\Sigma_g\times S^1$    & $L_x \times T^2$ & $L_y \times T^2$ & $L_z \times \Sigma_g$
                        & $2L_x + 2L_y + 2gL_z - 3g$ & $3g$
\end{tabular}
\end{center}
\caption{
A summary of the ground state degeneracy (GSD) of the X-cube model on various 3-manifolds with the foliations described in \secref{sec:manifolds}. *The logical operators with support near foliation singularities are not protected against local perturbations; see \secref{sec:S2S1}.}\label{tab:GSD}
\end{table*}

\subsection{\texorpdfstring{$\bf{S^2\times S^1}$}{S\^{}2 x S\^{}1}}
\label{sec:S2S1}

First, consider the manifold $S^2\times S^1$. It admits a non-singular foliation consisting of layered copies of $S^2$, as well as singular foliations of tori whose projections onto $S^2$ latitudinally foliate the sphere with singularities at the poles. Our construction takes one stack of $L_z$ parallel spheres and two such stacks of $L_x$ and $L_y$ layers of tori, respectively (see \sfigref{fig:manifolds}{a}). The ground state degeneracy of the X-cube model on this lattice obeys the formula $\log_2{\textrm{GSD}}=2L_x+2L_y-1$.

It is important to note that the Wilson loops (which are a product of $X$ operators around a red or green loop in \sfigref{fig:manifolds}{a}) near the foliation singularities have local support. Thus, the logical qubits corresponding to these loops are not topologically protected, and the ground state degeneracy would be partially split by local perturbations. 

\subsection{3-sphere}
\label{sec:S3}

The 3-sphere $S^3$ admits latitudinal foliations with polar singularities. Viewing $S^3$ as a subspace of $\mathbb{R}^4$ defined by the equation $x^2+y^2+z^2+w^2=1$, a leaf of an $x$-oriented latitudinal foliation is a 2-sphere defined by the equation $x_0^2+y^2+z^2+w^2=1$ for fixed $x_0$. Taking three such foliations in the $x$, $y$, and $z$ directions yields a suitable cellulation of $S^3$ (\sfigref{fig:manifolds}{b}). The resulting X-cube model exhibits a unique ground state.

\subsection{Half-twist manifold}
\label{sec:halfTwist}

The half-twist manifold is an orientable Euclidean 3-manifold constructed by identifying opposite faces of a cube. The $y$ (top and bottom) and $x$ (left and right) faces are identified in the standard way, but the $z$ (front and back) faces are identified after a rotation of 180 degrees relative to one another.
It admits a total foliation with three sets of compact toroidal leaves. 
A sampling of $L_x$, $L_y$, and $L_z$ toroidal leaves corresponds to embedding a $2L_x\times2L_y\times L_z$ cubic lattice in the original cube. 
The factors of 2 are due to the twist in the gluing process (see \sfigref{fig:manifolds}{c}).
The ground state degeneracy of the X-cube model defined on this lattice is given by $\log_2{\textrm{GSD}}=2L_x+2L_y+2L_z$. 

It is also possible to include one or more Klein bottles in the cellulation. Including one Klein bottle belonging to the $x$ foliation corresponds to embedding a cubic lattice of size $2L_x-1$ in the $x$ direction, and changes the ground state degeneracy such that $c=2$. Including an additional Klein bottle in the $y$ foliation further increases $c$ to 3. Thus the constant $c$ is not an invariant of $M$, but rather is sensitive to the choice of cellulation.

\subsection{Klein bottle times \texorpdfstring{$\bf{S^1}$}{S\^{}1}}

The manifold $K^2\times S^1$ is a simple example of a non-orientable 3-manifold, where $K^2$ is a Klein bottle. It admits a total foliation consisting of one set of Klein bottle leaves and two sets of toroidal leaves. Cellulating $K^2\times S^1$ in this fashion (with $L_x$, $L_y$, and $L_z$ leaves) is equivalent to embedding a $2L_x\times 2L_y\times L_z$ cubic lattice into $K^2\times S^1$. 
See \sfigref{fig:manifolds}{d} for details. As on the torus, the toric code on the Klein bottle has a fourfold ground state degeneracy. The ground state degeneracy of the X-cube model on $K^2\times S^1$ (with this foliation) obeys $\log_2{\textrm{GSD}}=2L_x+2L_y+2L_z-2$.

\subsection{Genus \texorpdfstring{$\bf\it{{g}}$}{g} surface times \texorpdfstring{$\bf{S^1}$}{S\^{}1}}

The product manifold $\Sigma_g\times S^1$ admits a natural foliation of $\Sigma_g$ leaves, where $\Sigma_g$ is the 2D oriented topological manifold with genus $g$. We can supplement this with two singular foliations of tori which intersect $\Sigma_g$ slices in circles. These circles represent leaves of a \textit{measured foliation} of $\Sigma_g$ with trivalent singularities, which can be constructed for any genus by gluing together foliated pairs of pants.\cite{Casson} A cross-section of a cellulation of $\Sigma_2\times S^1$ is shown in \sfigref{fig:manifolds}{e}. The ground state degeneracy of the X-cube model on this class of manifolds is given by $\log_2{\textrm{GSD}}=2gL_z+2L_x+2L_y-3g$.

\section{Entanglement renormalization for X-cube model}
\label{sec:RG}

In this section, we introduce a procedure that transforms between X-cube ground states on coarsened or refined cellulations of a 3-manifold $M$. Given a cellulation composed of leaves of a triple foliation of $M$, the procedure allows leaves to be added or removed by adding or removing toric code states that live on the individual layers. This transformation sheds light on the structure of the ground state degeneracy in \eqnref{eq:GSDgenus}.
Moreover, the procedure can be interpreted as an entanglement renormalization group (RG) transformation\cite{VidalRG} for which the X-cube Hamiltonian is a fixed point. This interpretation motivates us to propose a definition of \textit{type-I fracton phase} based on ground state entanglement structure (in \secref{sec:type1}).

\subsection{RG transformation}

To begin, we consider a transformation on an X-cube ground state $\ket{\psi_\textrm{XC}}$ that adds a single layer to one of the constituent stacks of a given 3-manifold cellulation.
This new layer bisects the edges and 3-cells it intersects.
Accordingly, for each edge $i$ piercing the new layer,
the qubit on this edge is split into two qubits on edges $i'$ and $j'$.
We then substitute $Z_i \rightarrow Z_{i'}$ and $X_i \rightarrow X_{i'} X_{j'}$ in the X-cube Hamiltonian, and add a $Z_{i'} Z_{j'}$ stabilizer to the Hamiltonian.
\footnote{More formally, the edge can be split by initializing the new qubit $j'$ in the state $\ket{0}$ and applying a CNOT with control $i'$ and target $j'$.}
The resulting ground state is labeled $\ket{\psi_\textrm{XC}}'$.
Next, we take the tensor product $\ket{\psi_\textrm{XC}}'\otimes\ket{\psi_{\textrm{TC}}}$ of the modified X-cube ground state with a toric code ground state on the new layer, and apply a local unitary transformation $S$ to sew the two wave functions together into a larger X-cube wave function $\widetilde{\ket{\psi_\textrm{XC}}}=S\left(\ket{\psi_\textrm{XC}}'\otimes\ket{\psi_{\textrm{TC}}}\right)$. This procedure can be reversed or iterated to arbitrarily change the system size.

\begin{figure}
    \centering
    \begin{tikzpicture}
        \pgfmathsetmacro{\l}{2.4}

        \draw[line width=.5]
        (-\l,-\l,0) --++ (0,0,-\l)
        (-\l,0,0) --++ (0,0,-\l)
        (0,-\l,0) --++ (0,0,-\l)
        (0,0,0) --++ (0,0,-\l)
        (0,0,0) -- ++(-\l,0,0) -- ++(0,-\l,0) -- ++(\l,0,0) -- cycle
        (0,0,-\l) -- ++(-\l,0,0) -- ++(0,-\l,0) -- ++(\l,0,0) -- cycle;
        
        \draw[line width=1,color={rgb:black,1;blue,1}]
        (-\l/2,0,0) -- ++(0,0,-\l)
        (-\l/2,-\l,0) -- ++(0,0,-\l)
        (-\l/2,-\l,-\l) -- ++(0,\l,0)
        (-\l/2,-\l,0) -- ++(0,\l,0);
        
        \draw[arrows={-latex}, thick] (-\l/2+.15,-\l/2,0) -- ++(\l/2-.25,0,0);
        \draw[arrows={-latex}, thick] (-\l/2+.15,0,-\l/2) -- ++(\l/2-.25,0,0);
        \draw[arrows={-latex}, thick] (-\l/2+.15,-\l/2,-\l) -- ++(\l/2-.25,0,0);
        \draw[arrows={-latex}, thick] (-\l/2+.15,-\l,-\l/2) -- ++(\l/2-.25,0,0);
        
        \draw[arrows={-latex}, thick] (-\l/2+.133,0,-\l/2+.1) -- (-\l/4,0,-.1);
        \draw[arrows={-latex}, thick] (-\l/2+.15,0,-\l/2-.15) -- ++(\l/4-.15,0,-\l/2+.25);
        
        \draw[arrows={-latex}, thick] (-\l/2+.133,-\l,-\l/2+.1) -- (-\l/4,-\l,-.1);
        \draw[arrows={-latex}, thick] (-\l/2+.15,-\l,-\l/2-.15) -- ++(\l/4-.15,0,-\l/2+.25);
        
        \draw[arrows={-latex}] (+\l/2,-\l,-\l) -- ++(\l/2,0,0);
        \draw[arrows={-latex}] (+\l/2,-\l,-\l) -- ++(0,\l/2,0);
        \draw[arrows={-latex}] (+\l/2,-\l,-\l) -- ++(0,0,\l/1.5);
        
        \draw (\l/2+.2,-\l+1.2,-\l) node[fill=none] {$x$};
        \draw (\l-.3,-\l-.2,-\l) node[fill=none] {$z$};
        \draw (\l/2-.7,-\l-.4,-\l) node[fill=none] {$y$};
        
        \draw(-\l/4*3,-\l*1.1,0) node[fill=none] {$i'$};
        \draw(-\l/4*2,-\l*1.1,0) node[fill=none] {$\alpha$};
        \draw(-\l/4,-\l*1.1,0) node[fill=none] {$j'$};
        \draw(-\l/4*0,-\l*1.1,0) node[fill=none] {$\beta$};
        
        \draw (-\l*1.1,\l*.3,0) node[fill=white] {\bf (a)};
        
        

    \end{tikzpicture}
    \\\vspace{.4cm}
    \begin{tikzpicture}
    \pgfmathsetmacro{\l}{3.5}
        \draw[color={rgb:green,.5;black,1}, line width=.5, double]
        (-\l*.5,0,0)--++(-\l*.5,0,-\l*.6)
        (-\l*.8,0,-\l*1.2)--++(\l*.5,0,0)
        (\l*.2,0,-\l*.6)--++(-\l*.2,0,\l*.6);
        \draw[color={rgb:black,1;blue,1}, line width=1]
        (0,0,0) --++ (-\l/2,0,0)
        (-\l*1,0,-\l*.6)--++ (\l*.2,0,-\l*.6)
        (-\l*.3,0,-\l*1.2)--++ (\l*.5,0,\l*.6);
        \draw[line width=.5]
        (0,\l*.3,0) --++ (-\l/2,0,0) --++ (-\l*.5,0,-\l*.6)--++ (\l*.2,0,-\l*.6) --++ (\l*.5,0,0) --++ (\l*.5,0,\l*.6)--++ (-\l*.2,0,\l*.6) -- cycle;
        \draw[line width=.5]
        (0,0,0)--++(0,\l*.3,0)
        (-\l*.5,0,0)--++(0,\l*.3,0)
        (-\l*1,0,-\l*.6)--++(0,\l*.3,0)
        (-\l*.8,0,-\l*1.2)--++(0,\l*.3,0)
        (-\l*.3,0,-\l*1.2)--++(0,\l*.3,0)
        (\l*.2,0,-\l*.6)--++(0,\l*.3,0);

        \draw[arrows={-latex}, thick](-\l*.25,\l*.02,0) --++ (0,\l*.26,0);
        \draw[arrows={-latex}, thick](-\l*.27,\l*.02,0) --++ (-\l*.21,\l*.13,0);
        \draw[arrows={-latex}, thick](-\l*.23,\l*.02,0) --++ (\l*.21,\l*.13,0);
        
        \draw[arrows={-latex}, thick](-\l*.75,\l*.02,-\l*.3) --++ (0,\l*.26,0);
        
        \draw[arrows={-latex}, thick](-\l*.9,\l*.02,-\l*.9) --++ (0,\l*.26,0);
        \draw[arrows={-latex}, thick](-\l*.895,\l*.02,-\l*.9) --++ (\l*.08,\l*.13,-\l*.3);
        \draw[arrows={-latex}, thick](-\l*.905,\l*.02,-\l*.9) --++ (-\l*.08,\l*.13,\l*.3);
        
        \draw[arrows={-latex}, thick](-\l*.55,\l*.02,-\l*1.2) --++ (0,\l*.26,0);
        
        \draw[arrows={-latex}, thick](-\l*.05,\l*.02,-\l*.9) --++ (0,\l*.26,0);
        \draw[arrows={-latex}, thick](-\l*.045,\l*.02,-\l*.9) --++ (\l*.23,\l*.13,\l*.3);
        \draw[arrows={-latex}, thick](-\l*.055,\l*.02,-\l*.9) --++ (-\l*.23,\l*.13,-\l*.3);
        
        \draw[arrows={-latex}, thick](\l*.1,\l*.02,-\l*.3) --++ (0,\l*.26,0);
        
        \draw (-\l*.9,\l*.05,0) node[fill=white] {\bf (b)};
        
        \draw[arrows={-latex}] (+\l*.75,.4,0) -- ++(0,1.2,0);
        \draw (\l*.75+.2,1.2,0) node[fill=none] {$z$};
        
    \end{tikzpicture}
    \caption{{\bf (a)} Adding an $xy$-layer to the X-cube model on $T^3$. The large cube represents a unit cell of the original X-cube model, while the bold (blue) square is an elementary plaquette of the new layer $\alpha$. The original $z$-oriented edges are split into two by the new layer. The local unitary $S$ is a translation-invariant composition of commuting CNOT gates; a unit cell is pictured here. Arrows point from control qubit to target qubit. {\bf (b)} Action of the unitary $S$ on the qubits of a hexagonal prism 3-cell. The lower hexagonal plaquette belongs to the new $z$ layer $\alpha$. Bold (blue) edges are transverse to the $y$ foliation, whereas double (green) edges are tranverse to the $x$ foliation.}
    \label{fig:renormalization}
\end{figure}
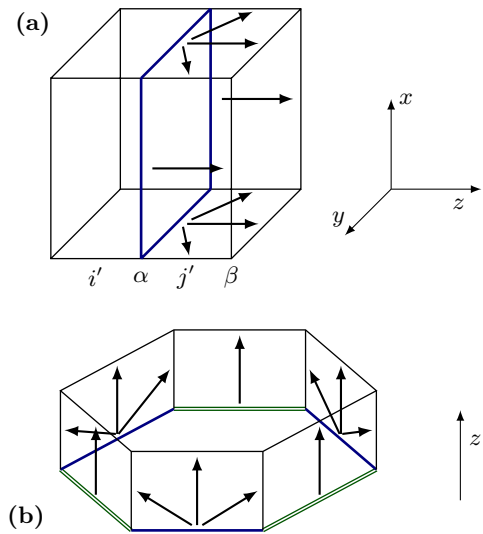

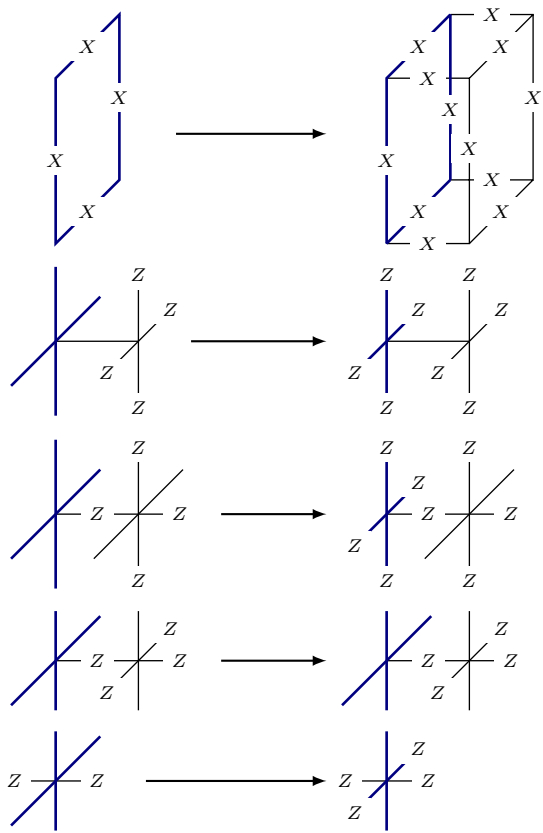
\begin{figure}
    \centering
    \begin{tikzpicture}
        \pgfmathsetmacro{\l}{2.2}
        \pgfmathsetmacro{\dif}{2*\l}
        \pgfmathsetmacro{\arr}{2.5}
        
        \draw[line width=1,color={rgb:black,1;blue,1}]
        (-\l/2,0,0) -- ++(0,0,-\l)--++(0,-\l,0)--++(0,0,\l)--cycle;
        
        \draw[line width=.5]
        (\dif-\l/2,0,0) -- ++(\l/2,0,0)--++(0,-\l,0) -- ++(-\l/2,0,0)
        (\dif-\l/2,0,-\l) -- ++(\l/2,0,0)--++(0,-\l,0) -- ++(-\l/2,0,0)
        (\dif,0,0)--++(0,0,-\l)
        (\dif,-\l,0)--++(0,0,-\l);
        
        \draw[line width=1,color={rgb:black,1;blue,1}]
        (\dif-\l/2,-\l,0) -- ++(0,0,-\l)
        (\dif-\l/2,-\l,-\l) -- ++(0,\l,0)
        (\dif-\l/2,-\l,0) -- ++(0,\l,0) --++(0,0,-\l);
        
        \draw (-\l/2,-\l/2,0) node[fill=white] {\scriptsize$X$};
        \draw (-\l/2,-\l/2,-\l) node[fill=white] {\scriptsize$X$};
        \draw (-\l/2,-\l,-\l/2) node[fill=white] {\scriptsize$X$};
        \draw (-\l/2,0,-\l/2) node[fill=white] {\scriptsize$X$};
        
        \draw (\dif-\l/2,-\l/2,0) node[fill=white] {\scriptsize$X$};
        \draw (\dif-\l/2,-\l*.575,-\l) node[fill=white] {\scriptsize$X$};
        \draw (\dif-\l/2,-\l,-\l/2) node[fill=white] {\scriptsize$X$};
        \draw (\dif-\l/2,0,-\l/2) node[fill=white] {\scriptsize$X$};
        \draw (\dif,-\l*.425,0) node[fill=white] {\scriptsize$X$};
        \draw (\dif,-\l/2,-\l) node[fill=white] {\scriptsize$X$};
        \draw (\dif,-\l,-\l/2) node[fill=white] {\scriptsize$X$};
        \draw (\dif,0,-\l/2) node[fill=white] {\scriptsize$X$};
        \draw (\dif-\l/4,0,0) node[fill=white] {\scriptsize$X$};
        \draw (\dif-\l/4,0,-\l) node[fill=white] {\scriptsize$X$};
        \draw (\dif-\l/4,-\l,0) node[fill=white] {\scriptsize$X$};
        \draw (\dif-\l/4,-\l,-\l) node[fill=white] {\scriptsize$X$};
        
        \draw[arrows={-latex},thick] (.5,-\l*.7+.8,0) -- (\arr,-\l*.7+.8,0);

        \pgfmathsetmacro{\di}{3.5}
        \draw [line width=.5](0,-\di,0) -- ++ (-\l/2,0,0);
        \draw [line width=.5](0,-\di,0) -- ++ (0,.4*\l,0);
        \draw [line width=.5](0,-\di,0) -- ++ (0,-.4*\l,0);
        \draw [line width=.5](0,-\di,0) -- ++ (0,0,.7*\l);
        \draw [line width=.5](0,-\di,0) -- ++ (0,0,-.7*\l);
        \draw [line width=1,color={rgb:black,1;blue,1}](-\l/2,-\di,0) -- ++ (0,.45*\l,0);
        \draw [line width=1,color={rgb:black,1;blue,1}](-\l/2,-\di,0) -- ++ (0,-.45*\l,0);
        \draw [line width=1,color={rgb:black,1;blue,1}](-\l/2,-\di,0) -- ++ (0,0,.7*\l);
        \draw [line width=1,color={rgb:black,1;blue,1}](-\l/2,-\di,0) -- ++ (0,0,-.7*\l);
        
        \draw [line width=.5](\dif,-\di,0) -- ++ (-\l/2,0,0);
        \draw [line width=.5](\dif,-\di,0) -- ++ (0,.4*\l,0);
        \draw [line width=.5](\dif,-\di,0) -- ++ (0,-.4*\l,0);
        \draw [line width=.5](\dif,-\di,0) -- ++ (0,0,.7*\l);
        \draw [line width=.5](\dif,-\di,0) -- ++ (0,0,-.7*\l);
        \draw [line width=1,color={rgb:black,1;blue,1}](\dif-\l/2,-\di,0) -- ++ (0,.4*\l,0);
        \draw [line width=1,color={rgb:black,1;blue,1}](\dif-\l/2,-\di,0) -- ++ (0,-.4*\l,0);
        \draw [line width=1,color={rgb:black,1;blue,1}](\dif-\l/2,-\di,0) -- ++ (0,0,.7*\l);
        \draw [line width=1,color={rgb:black,1;blue,1}](\dif-\l/2,-\di,0) -- ++ (0,0,-.7*\l);
        
        \draw (0,-\di-\l*.4,0) node[fill=white] {\scriptsize$Z$};
        \draw (0,-\di+\l*.4,0) node[fill=white] {\scriptsize$Z$};
        \draw (0,-\di,-\l*.5) node[fill=white] {\scriptsize$Z$};
        \draw (0,-\di,\l*.5) node[fill=white] {\scriptsize$Z$};
        
        \draw (\dif,-\di-\l*.4,0) node[fill=white] {\scriptsize$Z$};
        \draw (\dif,-\di+\l*.4,0) node[fill=white] {\scriptsize$Z$};
        \draw (\dif,-\di,-\l*.5) node[fill=white] {\scriptsize$Z$};
        \draw (\dif,-\di,\l*.5) node[fill=white] {\scriptsize$Z$};
        \draw (\dif-\l/2,-\di-\l*.4,0) node[fill=white] {\scriptsize$Z$};
        \draw (\dif-\l/2,-\di+\l*.4,0) node[fill=white] {\scriptsize$Z$};
        \draw (\dif-\l/2,-\di,-\l*.5) node[fill=white] {\scriptsize$Z$};
        \draw (\dif-\l/2,-\di,\l*.5) node[fill=white] {\scriptsize$Z$};
        
        \draw[arrows={-latex},thick] (.7,-\di,0) -- (\arr,-\di,0);

        \pgfmathsetmacro{\dii}{\di+2.3}
        \draw [line width=.5](-\l/2,-\dii,0) -- ++ (\l*.8,0,0);
        \draw [line width=.5](0,-\dii,0) -- ++ (0,.4*\l,0);
        \draw [line width=.5](0,-\dii,0) -- ++ (0,-.4*\l,0);
        \draw [line width=.5](0,-\dii,0) -- ++ (0,0,.7*\l);
        \draw [line width=.5](0,-\dii,0) -- ++ (0,0,-.7*\l);
        \draw [line width=1,color={rgb:black,1;blue,1}](-\l/2,-\dii,0) -- ++ (0,.45*\l,0);
        \draw [line width=1,color={rgb:black,1;blue,1}](-\l/2,-\dii,0) -- ++ (0,-.45*\l,0);
        \draw [line width=1,color={rgb:black,1;blue,1}](-\l/2,-\dii,0) -- ++ (0,0,.7*\l);
        \draw [line width=1,color={rgb:black,1;blue,1}](-\l/2,-\dii,0) -- ++ (0,0,-.7*\l);
        
        \draw [line width=.5](\dif-\l/2,-\dii,0) -- ++ (\l*.8,0,0);
        \draw [line width=.5](\dif,-\dii,0) -- ++ (0,.4*\l,0);
        \draw [line width=.5](\dif,-\dii,0) -- ++ (0,-.4*\l,0);
        \draw [line width=.5](\dif,-\dii,0) -- ++ (0,0,.7*\l);
        \draw [line width=.5](\dif,-\dii,0) -- ++ (0,0,-.7*\l);
        \draw [line width=1,color={rgb:black,1;blue,1}](\dif-\l/2,-\dii,0) -- ++ (0,.4*\l,0);
        \draw [line width=1,color={rgb:black,1;blue,1}](\dif-\l/2,-\dii,0) -- ++ (0,-.4*\l,0);
        \draw [line width=1,color={rgb:black,1;blue,1}](\dif-\l/2,-\dii,0) -- ++ (0,0,.7*\l);
        \draw [line width=1,color={rgb:black,1;blue,1}](\dif-\l/2,-\dii,0) -- ++ (0,0,-.7*\l);
        
        \draw (0,-\dii-\l*.4,0) node[fill=white] {\scriptsize$Z$};
        \draw (0,-\dii+\l*.4,0) node[fill=white] {\scriptsize$Z$};
        \draw (\l/4,-\dii,0) node[fill=white] {\scriptsize$Z$};
        \draw (-\l/4,-\dii,0) node[fill=white] {\scriptsize$Z$};
        
        \draw (\dif,-\dii-\l*.4,0) node[fill=white] {\scriptsize$Z$};
        \draw (\dif,-\dii+\l*.4,0) node[fill=white] {\scriptsize$Z$};
        \draw (\dif+\l/4,-\dii,0) node[fill=white] {\scriptsize$Z$};
        \draw (\dif-\l/4,-\dii,0) node[fill=white] {\scriptsize$Z$};
        \draw (\dif-\l/2,-\dii-\l*.4,0) node[fill=white] {\scriptsize$Z$};
        \draw (\dif-\l/2,-\dii+\l*.4,0) node[fill=white] {\scriptsize$Z$};
        \draw (\dif-\l/2,-\dii,-\l*.5) node[fill=white] {\scriptsize$Z$};
        \draw (\dif-\l/2,-\dii,\l*.5) node[fill=white] {\scriptsize$Z$};
        
        \draw[arrows={-latex}, thick] (1.1,-\dii,0) -- (\arr,-\dii,0);
        
        \pgfmathsetmacro{\diii}{\dii+1.95}
        \draw [line width=.5](-\l/2,-\diii,0) -- ++ (\l*.8,0,0);
        \draw [line width=.5](0,-\diii,0) -- ++ (0,.3*\l,0);
        \draw [line width=.5](0,-\diii,0) -- ++ (0,-.3*\l,0);
        \draw [line width=.5](0,-\diii,0) -- ++ (0,0,.7*\l);
        \draw [line width=.5](0,-\diii,0) -- ++ (0,0,-.7*\l);
        \draw [line width=1,color={rgb:black,1;blue,1}](-\l/2,-\diii,0) -- ++ (0,.3*\l,0);
        \draw [line width=1,color={rgb:black,1;blue,1}](-\l/2,-\diii,0) -- ++ (0,-.3*\l,0);
        \draw [line width=1,color={rgb:black,1;blue,1}](-\l/2,-\diii,0) -- ++ (0,0,.7*\l);
        \draw [line width=1,color={rgb:black,1;blue,1}](-\l/2,-\diii,0) -- ++ (0,0,-.7*\l);
        
        \draw [line width=.5](\dif-\l/2,-\diii,0) -- ++ (\l*.8,0,0);
        \draw [line width=.5](\dif,-\diii,0) -- ++ (0,.3*\l,0);
        \draw [line width=.5](\dif,-\diii,0) -- ++ (0,-.3*\l,0);
        \draw [line width=.5](\dif,-\diii,0) -- ++ (0,0,.7*\l);
        \draw [line width=.5](\dif,-\diii,0) -- ++ (0,0,-.7*\l);
        \draw [line width=1,color={rgb:black,1;blue,1}](\dif-\l/2,-\diii,0) -- ++ (0,.3*\l,0);
        \draw [line width=1,color={rgb:black,1;blue,1}](\dif-\l/2,-\diii,0) -- ++ (0,-.3*\l,0);
        \draw [line width=1,color={rgb:black,1;blue,1}](\dif-\l/2,-\diii,0) -- ++ (0,0,.7*\l);
        \draw [line width=1,color={rgb:black,1;blue,1}](\dif-\l/2,-\diii,0) -- ++ (0,0,-.7*\l);
        
        \draw (0,-\diii,-\l/2) node[fill=white] {\scriptsize$Z$};
        \draw (0,-\diii,\l/2) node[fill=white] {\scriptsize$Z$};
        \draw (\l/4,-\diii,0) node[fill=white] {\scriptsize$Z$};
        \draw (-\l/4,-\diii,0) node[fill=white] {\scriptsize$Z$};
        
        \draw (\dif,-\diii,-\l/2) node[fill=white] {\scriptsize$Z$};
        \draw (\dif,-\diii,\l/2) node[fill=white] {\scriptsize$Z$};
        \draw (\dif+\l/4,-\diii,0) node[fill=white] {\scriptsize$Z$};
        \draw (\dif-\l/4,-\diii,0) node[fill=white] {\scriptsize$Z$};
        
        \draw[arrows={-latex}, thick] (1.1,-\diii,0) -- (\arr,-\diii,0);
        
        \pgfmathsetmacro{\diiii}{\diii+1.6}
        \draw [line width=.5](-\l*.2,-\diiii,0) -- ++ (-\l*.6,0,0);
        \draw [line width=1,color={rgb:black,1;blue,1}](-\l/2,-\diiii,0) -- ++ (0,.3*\l,0);
        \draw [line width=1,color={rgb:black,1;blue,1}](-\l/2,-\diiii,0) -- ++ (0,-.3*\l,0);
        \draw [line width=1,color={rgb:black,1;blue,1}](-\l/2,-\diiii,0) -- ++ (0,0,.7*\l);
        \draw [line width=1,color={rgb:black,1;blue,1}](-\l/2,-\diiii,0) -- ++ (0,0,-.7*\l);
        
        \draw [line width=.5](\dif-\l*.2,-\diiii,0) -- ++ (-\l*.6,0,0);
        \draw [line width=1,color={rgb:black,1;blue,1}](\dif-\l/2,-\diiii,0) -- ++ (0,.3*\l,0);
        \draw [line width=1,color={rgb:black,1;blue,1}](\dif-\l/2,-\diiii,0) -- ++ (0,-.3*\l,0);
        \draw [line width=1,color={rgb:black,1;blue,1}](\dif-\l/2,-\diiii,0) -- ++ (0,0,.7*\l);
        \draw [line width=1,color={rgb:black,1;blue,1}](\dif-\l/2,-\diiii,0) -- ++ (0,0,-.7*\l);
        
        \draw (-\l/4,-\diiii,0) node[fill=white] {\scriptsize$Z$};
        \draw (-3*\l/4,-\diiii,0) node[fill=white] {\scriptsize$Z$};
        
        \draw (\dif-\l/4,-\diiii,0) node[fill=white] {\scriptsize$Z$};
        \draw (\dif-3*\l/4,-\diiii,0) node[fill=white] {\scriptsize$Z$};
        \draw (\dif-\l/2,-\diiii,-\l/2) node[fill=white] {\scriptsize$Z$};
        \draw (\dif-\l/2,-\diiii,\l/2) node[fill=white] {\scriptsize$Z$};
        
        \draw[arrows={-latex}, thick] (.1,-\diiii,0) -- (\arr,-\diiii,0);

    \end{tikzpicture}
    \caption{Adjoint action of $S$ on stabilizers of $\ket{\psi_\textrm{XC}}'\otimes\ket{\psi_\textrm{TC}}$. As in Fig. \ref{fig:renormalization}, bold (blue) lines correspond to edges of the new layer. Terms not pictured are unchanged.}
    \label{fig:stabilizers}
\end{figure}

The unitary $S$ is a product of CNOT gates whose control qubits belong to the new layer; the precise form depends on the geometry of the cellulation. In the simplest case, the intersection of the new layer $\alpha$ with the two transverse foliations is isomorphic to that of one of the adjacent layers $\beta$. Suppose $\alpha$ and $\beta$ are $z$ leaves. The region between $\alpha$ and $\beta$ is divided by the $x$ and $y$ foliations into 3-cell prisms whose base polygons have an even number of sides alternating between edges transverse to the $x$ and $y$ foliations. The unitary $S$ contains a CNOT gate for each edge (the control qubit) in $\alpha$, whose target qubit lies on the corresponding edge in $\beta$. Additionally, for each edge transverse to the $y$ foliation in $\alpha$, there are two additional CNOT gates whose targets are the adjacent edges transverse to $\alpha$ (oriented in the $z$ direction) and connected to $\beta$. The transformation for a cubic lattice is illustrated in \sfigref{fig:renormalization}{a}, and for the edges in a hexagonal prism 3-cell in \hyperref[fig:renormalization]{(b)}. CNOT acts by conjugation as:
\begin{equation}
    \begin{split}
    ZI\to ZI \qquad IZ\leftrightarrow ZZ \\
    XI\leftrightarrow XX \qquad IX \to IX,
    \end{split}
\end{equation}
where the first and second qubits are the control and target qubits, respectively.
From this follows the action of $S$ on the generators of the stabilizer group of $\ket{\psi_\textrm{XC}}'\otimes\ket{\psi_{\textrm{TC}}}$, i.e. the modified X-cube Hamiltonian terms combined with toric code Hamiltonian terms on the new layer.
This action is described for a cubic lattice in \figref{fig:stabilizers}. The prism geometry ensures that $S$ maps the original stabilizer generators to a set of stabilizer generators corresponding to a larger X-cube model. It follows that $\widetilde{\ket{\psi_\textrm{XC}}}$ is indeed an X-cube ground state on the enlarged lattice.

In general, the leaves adjacent to the new layer $\alpha$ may not have isomorphic intersections with the other foliations, in which case the local unitary $S$ which sews $\alpha$ into the cellulation may take a complicated form. However, we believe that such an operator generically exists. In \appref{sec:A} we present examples of explicit transformations to add generic leaves to the total foliations of $S^2\times S^1$ and $S^3$ discussed in \secref{sec:manifolds}. For the other manifolds discussed, the even-faced prism construction is sufficient to freely change the system size.

\subsection{Ground state degeneracy}

Because the toric code ground space is degenerate (on topologically non-trivial surfaces), a branching structure is present in the renormalization process. For every surface added to a cellulation, there are $2^b$ possible toric code ground states which can be used as inputs for the RG transformation (where $b$ is the $\mathbb{Z}_2$ Betti number of the added surface); each of these choices corresponds to a different sector of X-cube ground states. Thus, the entanglement renormalization picture naturally explains the subextensive growth of the X-cube ground space described by \eqnref{eq:GSDtorus} and (\ref{eq:GSDgenus}): the scaling of the ground space degeneracy on a manifold $M$ arises from the non-trivial homology of the leaves that foliate $M$.

The constant $c$ in \eqnref{eq:GSDgenus} can be understood by considering the minimal cellulation (e.g. $L_x=L_y=L_z=1$) and viewing larger systems as entanglement RG outgrowths of this seed system. (For the case of $\Sigma_g\times S^1$ and cellulations of the half-twist manifold that contain at least one Klein bottle, the minimal cellulation obtainable from disentangling toric code layers contains more than 3 leaves.)
The X-cube Hilbert space can be viewed as the physical subspace of an extended Hilbert space which is a tensor product of toric code Hilbert spaces on each leaf, corresponding to two qubits per edge.
X-cube ground states can be written as $\ket{\psi_\mathrm{XC}}=\prod_e P_e\bigotimes_\ell\ket{\psi^\ell_\mathrm{TC}}$, where $\ell$ runs over leaves, $e$ runs over edges, $\ket{\psi^\ell_\mathrm{TC}}$ is a toric code ground state on leaf $\ell$, and $P_e=1+Z_iZ_j$ where $i$ and $j$ are the two qubits on edge $e$. The product $\prod_e P_e$ projects onto the physical subspace, and maps products of coinciding Wilson loops to the identity operator.
Thus, for the seed system, $c$ counts the redundancies in logical qubits of the minimal leaves, or in other words, the number of leaf intersections which correspond to non-trivial first homology classes of both leaves. Thus, $c$ is sensitive to the foliating structure on $M$, and in particular to the way the foliations intersect.
These considerations can be used to compute the values of $c$ shown in Table \ref{tab:GSD} and explain the dependence of $c$ for the half-twist manifold on the presence or absence of Klein bottles in the cellulation, as discussed in \secref{sec:halfTwist}.

\subsection{Relation to the Haah code}

The RG transformation presented here is related to a similar transformation for the Haah code studied in \refcite{HaahRG}. The Haah code is a type-II fracton model defined on a torus with non-trivial ground-state degeneracy, fractal-like excitation structure, and no string-like logical operators. The procedure of \refcite{HaahRG} employs a local unitary transformation $U$ to decouple the Haah code Hamiltonian $H_A$ on a cubic lattice of size $2L$  into two Hamiltonians $H_A$ and $H_B$ acting separately on interlacing sublattices of size $L$. Similar to the Haah code, $H_B$ is a type-II fracton model with fractal-like excitations. Conversely however, $H_B$ admits an RG transformation in which the model on a lattice of size $2L$ is related via a local unitary transformation $V$ to two copies of itself on interwoven size $L$ sublattices. This information is summarized as follows:
\begin{align}
   UH_A(2L)U^\dagger &\cong H_A(L)+H_B(L) \label{eq:RGbranch}\\
  VH_B(2L)V^\dagger &\cong H_B(L)+H_B(L), \nn
  \label{eq:RGbranch}
\end{align}
where $H\cong H'$ implies that $H$ and $H'$ have coinciding ground spaces corresponding to identical stabilizer groups. The X-cube RG transformation can be cast in the same light: $H_A$ is the X-cube Hamiltonian, whereas $H_B$ corresponds to three mutually perpendicular decoupled stacks of toric codes. We note that the branching structure of \eqnref{eq:RGbranch} indicates that X-cube ground states bear exact representations as branching MERA tensor networks.\cite{BranchingRG,Evenbly14}

\subsection{Entanglement structure}
Moreover, the existence of this RG transformation underlies the entanglement structure of the X-cube ground states. Because local unitary transformations do not modify the long-range entanglement structure, the entanglement entropy of a region $R$ can be heuristically understood as a combination of contributions from underlying toric code layers. The subleading linear correction to entanglement entropy for the X-cube model thus corresponds to a combination of the constant topological corrections present in toric code ground states. \cite{EntropyKP,EntropyLW} 
Interestingly, the Haah code also exhibits subleading linear corrections to entanglement entropy.\cite{HermeleEntropy,BernevigEntropy} Whether these corrections for the Haah code can be similarly understood from the entanglement RG perspective is not clear.

\section{Type-I fracton phases}
\label{sec:type1}

A paradigmatic understanding of 2D quantum phases in the absence of symmetry was reached by the authors of \refcite{Xie10}. In this framework, quantum phases are characterized by the pattern of long-range entanglement exhibited by their ground states, and correspond to unique 2D topological orders.\cite{Wen16} Two ground states are considered to represent the same quantum phase of matter if they are related by a generalized local unitary (gLU) transformation, which is a finite-depth quantum circuit augmented with free addition or removal of product states. System size can thus be altered by adding or removing product states and performing an appropriate local unitary transformation. In this sense, unentangled product states can be viewed as free `resources' for 2D quantum phases. However, in 3D, the gLU paradigm is unsatisfactory because it over-refines the space of ground states. While conventional 3D topological orders such as discrete gauge theories represent gLU equivalence classes, exotic fracton models such as the X-cube model and the Haah code (along with simple decoupled stacks of 2D topological orders) do not represent unique equivalence classes because ground states of different system sizes are not gLU-equivalent. The gLU framework is hence inadequate in 3D as it does not allow for a notion of thermodynamic limit.

For this reason we are motivated to propose a definition of \textit{type-I fracton phase} which incorporates the RG perspective of the X-cube model. In particular, we define a type-I fracton phase as a class of models exhibiting a thermodynamic limit whose ground state manifolds can be transformed into one another via tensor product with an arbitrary number of 2D topological ground states followed by the action of a finite-depth quantum circuit. In other words, we consider 2D topological orders as free resources for 3D fracton phases. In this sense, the X-cube model is a zero-correlation length fixed-point Hamiltonian under the entanglement RG transformation, and a representative model of a type-I fracton phase.

We note that the definition proposed here classifies decoupled stacks of 2D topological phases as trivial 3D phases. Moreover, it unifies the notions of type-I fracton order and conventional 3D topological order, where product state resources may be viewed as trivial 2D topological orders.
The definition we propose is closely related to and inspired by the $s$-sourcery framework introduced in \refcite{SwingleSSource}, which employs a more general notion of `resource' state and proposes a classification of all long-range entangled 3D quantum matter. The X-cube model provides a new example of a phase with matrix-valued $s$.



\section{\texorpdfstring{$\bf{\mathbb{Z}_N}$}{Z\_N} generalization of X-cube model}
\label{sec:ZN}
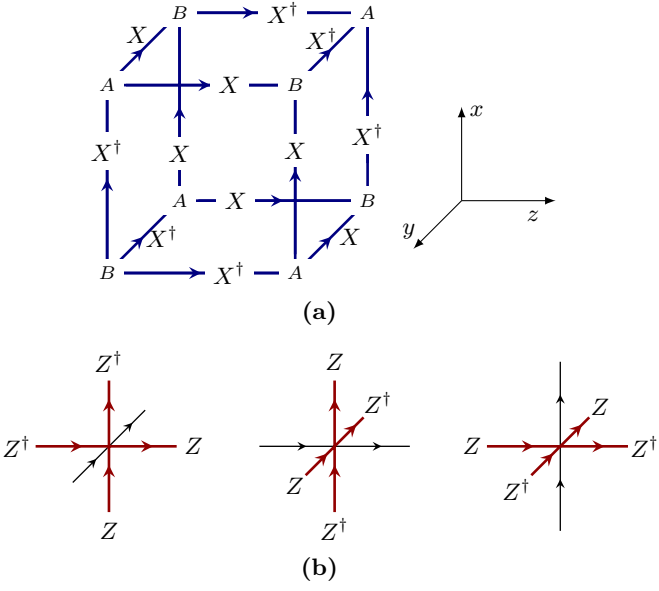
\begin{figure}

\tikzset{
  on each segment/.style={
    decorate,
    decoration={
      show path construction,
      moveto code={},
      lineto code={
        \path [#1]
        (\tikzinputsegmentfirst) -- (\tikzinputsegmentlast);
      },
      curveto code={
        \path [#1] (\tikzinputsegmentfirst)
        .. controls
        (\tikzinputsegmentsupporta) and (\tikzinputsegmentsupportb)
        ..
        (\tikzinputsegmentlast);
      },
      closepath code={
        \path [#1]
        (\tikzinputsegmentfirst) -- (\tikzinputsegmentlast);
      },
    },
  },
  mid arrow/.style={postaction={decorate,decoration={
        markings,
        mark=at position .5 with {\arrow[#1]{stealth}}
      }}},
}

    \begin{tikzpicture}
    
        \pgfmathsetmacro{\l}{2.5}

        \draw[line width=1,color={rgb:black,1;blue,1}]
        (0,0,0)--++(-\l,0,0)--++(0,-\l,0)--++(\l,0,0)--cycle
        (0,0,-\l)--++(-\l,0,0)--++(0,-\l,0)--++(\l,0,0)--cycle;

        \path [draw={rgb:black,1;blue,1}, line width=1,postaction={on each segment={mid arrow={}}}]
        (-\l,-\l,0)--++(\l,0,0)
        (-\l*.9,0,0)--++(\l*.9,0,0)
        (-\l*.9,-\l,-\l)--++(\l*.9,0,0)
        (-\l,0,-\l)--++(\l*.8,0,0)
        (0,-\l*.9,0)--++(0,\l*.9,0)
        (-\l,-\l,-\l)--++(0,\l,0)
        (-\l,-\l,0)--++(0,\l,0)
        (0,-\l*.8,-\l)--++(0,\l*.8,0)
        (0,0,0)--++(0,0,-\l)
        (-\l,0,0)--++(0,0,-\l)
        (0,-\l,0)--++(0,0,-\l)
        (-\l,-\l,0)--++(0,0,-\l);
        
        \draw (-\l*.35,-\l,0) node[fill=white] {$X^\dagger$};
        \draw (-\l*.7,-\l,-\l) node[fill=white] {$X$};
        \draw (-\l*.45,0,-\l) node[fill=white] {$X^\dagger$};
        \draw (-\l*.35,0,0) node[fill=white] {$X$};
        \draw (0,-\l*.35,0) node[fill=white] {$X$};
        \draw (-\l,-\l*.35,0) node[fill=white] {$X^\dagger$};
        \draw (-\l,-\l*.75,-\l) node[fill=white] {$X$};
        \draw (0,-\l*.65,-\l) node[fill=white] {$X^\dagger$};
        \draw (-\l*1.11,0,-\l*.7) node[fill=none] {$X$};
        \draw (-\l*0.11,0,-\l*.65) node[fill=none] {$X^\dagger$};
        \draw (-\l*.9,-\l,-\l*.5) node[fill=none] {$X^\dagger$};
        \draw (\l*.1,-\l,-\l*.5) node[fill=none] {$X$};
        
        \draw (0,0,0) node[fill=white] {\scriptsize $B$};
        \draw (-\l,0,0) node[fill=white] {\scriptsize $A$};
        \draw (0,-\l,0) node[fill=white] {\scriptsize $A$};
        \draw (-\l,-\l,0) node[fill=white] {\scriptsize $B$};
        \draw (0,0,-\l) node[fill=white] {\scriptsize $A$};
        \draw (-\l,0,-\l) node[fill=white] {\scriptsize $B$};
        \draw (0,-\l,-\l) node[fill=white] {\scriptsize $B$};
        \draw (-\l,-\l,-\l) node[fill=white] {\scriptsize $A$};
        
        \draw[arrows={-latex}] (+\l/2,-\l,-\l) -- ++(\l/2,0,0);
        \draw[arrows={-latex}] (+\l/2,-\l,-\l) -- ++(0,\l/2,0);
        \draw[arrows={-latex}] (+\l/2,-\l,-\l) -- ++(0,0,\l/1.5);
        
        \draw (\l/2+.2,-\l+1.2,-\l) node[fill=none] {$x$};
        \draw (\l-.3,-\l-.2,-\l) node[fill=none] {$z$};
        \draw (\l/2-.7,-\l-.4,-\l) node[fill=none] {$y$};
        
    \end{tikzpicture} \\
    {\bf (a)}\\\vspace{.2cm}
    \begin{tikzpicture}
        
        \pgfmathsetmacro{\l}{2.5}
        
        \pgfmathsetmacro{\ri}{0}
        
        \path[draw=black,line width=.5,postaction={on each segment={mid arrow={}}}] (.45*\l+\ri,-\l/2,0)--++(0,0,-\l*.65)--++(0,0,-\l*.35);
        \path[line width=1,draw={rgb:black,.75;red,1},postaction={on each segment={mid arrow={}}}]
        (\ri+.15,-\l/2,-\l/2) --++ (\l*.5,0,0) --++ (\l*.25,0,0)
        (.45*\l+\ri,-\l*.85,-\l/2) --++(0,\l*.5,0)--++(0,\l*.2,0);

        \pgfmathsetmacro{\rii}{\ri+1.45*\l}
        \path[draw=black,line width=.5,postaction={on each segment={mid arrow={}}}]
        (\rii-\l*.2,-\l/2,-\l/2) -- ++(\l*.5,0,0)-- ++(\l*.3,0,0);
        \path[line width=1,draw={rgb:black,.75;red,1},postaction={on each segment={mid arrow={}}}]
        (.2*\l+\rii,-\l/2,-\l*.1) --++ (0,0,-\l*.6)--++ (0,0,-\l*.2)
        (.2*\l+\rii,-\l*.85,-\l/2) --++(0,\l*.5,0)--++ (0,\l*.2,0);
        
        \pgfmathsetmacro{\riii}{\rii+.95*\l}
        \path[draw=black,line width=.5,,postaction={on each segment={mid arrow={}}}]
        (.45*\l+\riii,-\l*.95,-\l/2) -- ++(0,.55*\l,0)--++(0,.35*\l,0);
        \path[line width=1,draw={rgb:black,.75;red,1},postaction={on each segment={mid arrow={}}}]
        (.45*\l+\riii,-\l/2,-\l*.1) --++ (0,0,-\l*.6)--++ (0,0,-\l*.2)
        (\riii+.15,-\l/2,-\l/2) -- ++(\l*.5,0,0)--++(\l*.25,0,0);
        
        \draw (\ri-.1,-\l/2,-\l/2) node[fill=none] {$Z^\dagger$};
        \draw (\ri+.9*\l,-\l/2,-\l/2) node[fill=none] {$Z$};
        \draw (\ri+.45*\l,-\l/2+.45*\l,-\l/2) node[fill=none] {$Z^\dagger$};
        \draw (\ri+.45*\l,-\l/2-.45*\l,-\l/2) node[fill=none] {$Z$};
        
        \draw (\riii-.05,-\l/2,-\l/2) node[fill=none] {$Z$};
        \draw (\riii+.9*\l,-\l/2,-\l/2) node[fill=none] {$Z^\dagger$};
        \draw (\riii+.45*\l,-\l/2,-\l/2+.6*\l) node[fill=none] {$Z^\dagger$};
        \draw (\riii+.45*\l,-\l/2,-\l/2-.55*\l) node[fill=none] {$Z$};
        
        \draw (\rii+.2*\l,-\l/2+.45*\l,-\l/2) node[fill=none] {$Z$};
        \draw (\rii+.2*\l,-\l/2-.44*\l,-\l/2) node[fill=none] {$Z^\dagger$};
        \draw (\rii+.2*\l,-\l/2,-\l/2+.55*\l) node[fill=none] {$Z$};
        \draw (\rii+.2*\l,-\l/2,-\l/2-.6*\l) node[fill=none] {$Z^\dagger$};

    \end{tikzpicture} \\
    {\bf (b)}
    \caption{{\bf (a)} Action of 3-cell operator $B_c$ in $\mathbb{Z}_N$ X-cube Hamiltonian on a cubic 3-cell. Vertices of the cube have been given an A-B bipartition. {\bf (b)} Cross-shaped operators $A^\mu_v$ of the $\mathbb{Z}_N$ X-cube model Hamiltonian.}
    \label{fig:ZNstabilizers}
\end{figure}

The $\mathbb{Z}_N$ version of the X-cube model, as first discussed in \refcite{Slagle17}, is defined using the generalized Pauli operators $Z\ket{p}=\omega^p\ket{p}$ and $X\ket{p}=\ket{p+1\mod N}$, which act on dimension-$N$ local Hilbert spaces on each edge and obey the relations $ZX=\omega XZ$ and $Z^\dagger X=\omega^{-1}XZ^\dagger$ where $\omega=e^{2\pi i /N}$. In this section, we extend the $\mathbb{Z}_N$ model to general 3-manifolds cellulated by sets of transversely intersecting foliations, as in \secref{sec:manifolds}. To define the model it is necessary to orient each edge; reversing the orientation of an edge corresponds to inversion in $\mathbb{Z}_N$, given by $Z\leftrightarrow Z^\dagger$ and $X\leftrightarrow X^\dagger$. The Hamiltonian on any compact 3-manifold $M$ takes the form
\begin{equation}
    H=-\sum_v \left(A_v^x+A_v^y+A_v^z+\textrm{h.c.}\right) -\sum_c \left(B_c+B_c^\dagger\right).
\end{equation}
As in the $\mathbb{Z}_2$ case, $A_v^\mu$ is a cross-shaped operator at vertex $v$ whereas $B_c$ is a product of operators over the edges of the 3-cell $c$. The action of $A^\mu_v$ on an edge adjacent to $v$ is determined by the orientation and direction ($x$, $y$ or $z$) of the edge. $A^x_v$ acts as $Z$ ($Z^\dagger$) on $z$-directed ($y$-directed) edges whose orientations point towards $v$, and as $Z^\dagger$ ($Z$) on $y$-directed ($z$-directed) edges whose orientations point away from $v$, and likewise for cyclic permutations of $x$, $y$, and $z$. This is shown in \sfigref{fig:ZNstabilizers}{a} for a particular choice of orientations. On the other hand, to define the 3-cell term $B_c$, the vertices of $c$ are first given an A-B bipartition. A given 3-cell $c$ is guaranteed to be bipartite as a graph since all faces of $c$ have edges which sequentially alternate between two directions $\mu$ and $\nu$ (due to the foliating structure of the cellulation). $B_c$ is defined to act as $X$ on edges oriented from A to B vertices, and as $X^\dagger$ on edges oriented from B to A vertices (see \sfigref{fig:ZNstabilizers}{b}). The Hamiltonian terms mutually commute and constitute stabilizer generators for a dimension-$N$ qudit stabilizer code.

\begin{figure}

\tikzset{
  on each segment/.style={
    decorate,
    decoration={
      show path construction,
      moveto code={},
      lineto code={
        \path [#1]
        (\tikzinputsegmentfirst) -- (\tikzinputsegmentlast);
      },
      curveto code={
        \path [#1] (\tikzinputsegmentfirst)
        .. controls
        (\tikzinputsegmentsupporta) and (\tikzinputsegmentsupportb)
        ..
        (\tikzinputsegmentlast);
      },
      closepath code={
        \path [#1]
        (\tikzinputsegmentfirst) -- (\tikzinputsegmentlast);
      },
    },
  },
  mid arrow/.style={postaction={decorate,decoration={
        markings,
        mark=at position .5 with {\arrow[#1]{stealth}}
      }}},
}

    \centering
    \begin{tikzpicture}
        \pgfmathsetmacro{\l}{2.5}

        \path[draw=black,line width=.5,,postaction={on each segment={mid arrow={}}}]
        (-\l,-\l,0) --++ (0,0,-\l)
        (-\l,0,0) --++ (0,0,-\l)
        (0,-\l,0) --++ (0,0,-\l)
        (0,0,0) --++ (0,0,-\l)
        (-\l/2,-\l,0) --++ (0,0,-\l)
        (-\l/2,0,0) --++ (0,0,-\l)
        (-\l,0,0)--++(\l/2,0,0)
        (-\l,-\l,-\l)--++(\l/2,0,0)
        (-\l,0,-\l)--++(\l/2,0,0)
        (-\l,-\l,0)--++(\l/2,0,0)
        (-\l/2,0,0)--++(\l/2,0,0)
        (-\l/2,-\l,-\l)--++(\l/2,0,0)
        (-\l/2,0,-\l)--++(\l/2,0,0)
        (-\l/2,-\l,0)--++(\l/2,0,0)
        (-\l,-\l,0)--++(0,\l,0)
        (0,-\l,0)--++(0,\l,0)
        (-\l,-\l,-\l)--++(0,\l,0)
        (0,-\l,-\l)--++(0,\l,0)
        (-\l/2,-\l,0)--++(0,\l,0)
        (-\l/2,-\l,-\l)--++(0,\l,0)
        ;
        \draw[line width=.5]
        (0,0,0) -- ++(-\l,0,0) -- ++(0,-\l,0) -- ++(\l,0,0) -- cycle
        (0,0,-\l) -- ++(-\l,0,0) -- ++(0,-\l,0) -- ++(\l,0,0) -- cycle;
        
        \draw[line width=.8,color={rgb:black,1;blue,1}]
        (-\l/2,0,0) -- ++(0,0,-\l)
        (-\l/2,-\l,0) -- ++(0,0,-\l)
        (-\l/2,-\l,-\l) -- ++(0,\l,0)
        (-\l/2,-\l,0) -- ++(0,\l,0);
        
        \draw[arrows={-latex}, thick, double] (-\l/2+.15,-\l/2,0) -- ++(\l/2-.2,0,0);
        \draw[arrows={-latex}, thick, double] (-\l/2+.15,0,-\l/2) -- ++(\l/2-.25,0,0);
        \draw[arrows={-latex}, thick, double] (-\l/2+.15,-\l/2,-\l) -- ++(\l/2-.25,0,0);
        \draw[arrows={-latex}, thick, double] (-\l/2+.15,-\l,-\l/2) -- ++(\l/2-.25,0,0);
        
        \draw[arrows={-latex},thick] (-\l/2+.133,0,-\l/2+.1) -- (-\l/4,0,-.1);
        \draw[arrows={-latex}, thick, double] (-\l/2+.15,0,-\l/2-.2) -- ++(\l/4-.15,0,-\l/2+.35);
        
        \draw[arrows={-latex},thick] (-\l/2+.133,-\l,-\l/2+.1) -- (-\l/4,-\l,-.1);
        \draw[arrows={-latex}, thick, double] (-\l/2+.15,-\l,-\l/2-.2) -- ++(\l/4-.15,0,-\l/2+.35);
        

    \end{tikzpicture}
    \caption{Adding a layer to the $\mathbb{Z}_N$ X-cube model on a torus, as in Fig. \ref{fig:renormalization}. For the $\mathbb{Z}_N$ case, $S$ is a translation-invariant product of commuting $C$ and $C^\dagger$ operators; shown here is a unit cell. Arrows point from control qudit to target qudit; a single shaft indicates $C$ whereas a double shaft corresponds to $C^\dagger$.}
    \label{fig:ZNrenormalization}
\end{figure}
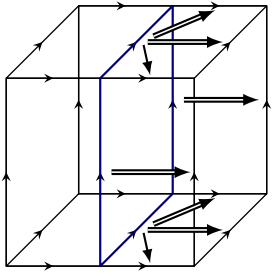

The physics of the $\mathbb{Z}_2$ model generalizes in a straightforward fashion to the $\mathbb{Z}_N$ setting, in which there are $N$ species of string and membrane operators obeying respective $\mathbb{Z}_N$ fusion rules. For prime $N$, the ground state degeneracy behaves identically, except that logical qubits are replaced with dimension-$N$ logical qudits. In particular, \eqnref{eq:GSDgenus} generalizes to the rule
\begin{equation}
    \log_N{\textrm{GSD}}=b_xL_x+b_yL_y+b_zL_z-c
\end{equation}
where $b_\mu$ is the first Betti number with $\mathbb{Z}_N$ coefficients. \cite{foot:Betti}
For composite (non-prime) $N$, the formula for ground state degeneracy is more complicated in general,\cite{Shinsei} since the ground space is not necessarily a tensor product of logical qudit Hilbert spaces.

In general, the scaling of the GSD can be understood in terms of an entanglement RG transformation which generalizes the discussion of \secref{sec:RG}.
For the $\mathbb{Z}_N$ X-cube model, generalized $\mathbb{Z}_N$ toric code states serve as two-dimensional resource states in the procedure (each contributing $b$ logical qudits for $N$ prime). Such transformations exist for all of the foliations we have discussed. As in the $\mathbb{Z}_2$ case, to add a layer we first split the qudits on edges $i$ intersecting the new layer into pairs of qudits $i'$ and $j'$, and add $Z_{i'}^\dagger Z_{j'}$ and $Z_{i'}Z_{j'}^\dagger$ stabilizer terms to the Hamiltonian and modify it as $Z_i\to Z_{i'}$ and $X_i\to X_{i'}X_{j'}$.
We then take the tensor product of the resulting $\mathbb{Z}_N$ X-cube state with a $\mathbb{Z}_N$ toric code state on the new layer, and apply a local unitary $S$. The operator $S$ is constructed from 2-qudit gates $C$ and $C^\dagger$ (see Fig. \ref{fig:ZNrenormalization} for the cubic lattice case), which are generalizations of the CNOT gate and act as $C\ket{p,q}=\ket{p,q+p}$ and $C^\dagger\ket{p,q}=\ket{p,q-p}$. The adjoint action of $C$ is given by 
\begin{equation}
    \begin{split}
    ZI\to ZI \qquad IZ\to Z^\dagger Z\\
    XI \to XX \qquad IX\to IX
    \end{split}
\end{equation}
whereas for $C^\dagger$ by
\begin{equation}
    \begin{split}
    ZI\to ZI \qquad IZ\to Z Z\\
    XI \to XX^\dagger \qquad IX\to IX.
    \end{split}
\end{equation}
It can be checked that $S$ maps the tensor product state to an enlarged $\mathbb{Z}_N$ X-cube ground state.

\section{Discussion}
\label{sec:discussion}

Our work on the $X$-cube model suggests that fracton physics could be regarded as a new kind of topological physics generalizing the traditional liquid topological order.\cite{ZengLiquid,QIQMbook} We conjecture the existence of the following $X$-cube TQFT.

A singular compact total foliation (SCTF) $\Lambda$ of a $3$-manifold $M$ consists of singular subsets $K_\mu$, $\mu=x,y,z$ (possibly empty) and three transversely intersecting sets of closed surfaces $\{\Lambda_x,\Lambda_y,\Lambda_z\}$ foliating the respective complements $M\backslash K_\mu$.
$K$ consists of singular leaves that are either finitely many points, a link,
\footnote{A link is an embedding of a finite number of circles into a 3-manifold, which may not intersect but may be linked or knotted.}
or some $G\times S^1$ where $G$ is a trivalent graph. Two SCTFs $\Lambda^{(1)}$ and $\Lambda^{(2)}$ on $M$ are considered to be equivalent if there exists a diffeomorphism $f$ of $M$ that sends $\Lambda^{(1)}$ to $\Lambda^{(2)}$ compatible with the singular leaves and the RG moves that define the fracton phase.
We believe every orientable closed $3$-manifold $M$ has an SCTF.
Given an SCTF $\Lambda$ on a $3$-manifold $M$, a finite $(L_x,L_y,L_z)$-version of $\Lambda$ is a choice of $L_x$, $L_y$, and  $L_z$ many leaves from the three stacks $\{\Lambda_x, \Lambda_y, \Lambda_z\}$, respectively, where $L_x$,$L_y$, and $L_z$ are natural numbers. \footnote{One could also consider 3+1D models with more or fewer than three sets of leaves.}

An SCTF-TQFT will assign to each pair $(M,\Lambda)$, where $\Lambda$ is an SCTF on the three manifold $M$, an infinite-dimensional Hilbert space $V(M,\Lambda)$ that is constructed as the limit of a sequence of finite versions of $\Lambda$.  Moreover, the GSD on the finite version $(L_x,L_y,L_z)$ depends only on the topology of $M$, the topology of the leaf surfaces, and the topology of the intersections of the leaves. The collection of Hilbert spaces $V(M,\Lambda)$ should satisfy some generalization of the usual TQFT axioms, and $V(M,\Lambda)$ is a representation of all diffeomorphisms of $M$ that preserve the SCTF $\Lambda$.  We will leave the construction of such an SCTF-TQFT for the $X$-cube model to the future.

As comparison, the authors of \refcite{Slagle17a} advocate that fracton models should be regarded as representing geometric orders.
Their approach was to consider how lattice geometry affects the low-energy physics and phase of matter as defined by generalized local unitary (gLU) equivalence.



It would be interesting to understand which components of this discussion generalize to other fracton models. For some of the type-I fracton models, a similar RG procedure can be identified; thus the SCTF structure may apply to these fracton models as well. On the other hand, type-II fracton models such as the Haah code do not fall within this framework. Moreover there is a class of gapless $U(1)$ fracton models. \cite{PretkoU1,electromagnetismPretko,Rasmussen2016,Xu2006,PretkoTheta,PretkoDuality}. It would be interesting to identify a substitute for the SCTF structure on general $3$-manifolds for these related models.

\begin{acknowledgments}
We are indebted to Michael Freedman, Ni Yi, Michael Pretko, Burak \c{S}ahino\v{g}lu, and Yong Baek Kim for inspiring discussions. We would also like to thank the Kavli Institute for Theoretical Physics where some of the discussion took place.
This research was supported in part by the National Science Foundation under Grant No. NSF PHY-1125915.
W. S. and X.C. are supported by thr National Science Foundation under award number DMR-1654340, the Alfred P. Sloan research fellowship, the Walter Burke Institute for Theoretical Physics, and the Institute for Quantum Information and Matter.
K.S. is supported by the NSERC of Canada and the Center for Quantum Materials at the University of Toronto.
Z.W. is supported by the National Science Foundation under award number DMR-1411212.

\end{acknowledgments}

\appendix

\section{Entanglement renormalization for 3-sphere and \texorpdfstring{$\bf{S^2\times S^1}$}{S\^{}2 x S\^{}1}}
\label{sec:A}

\begin{figure}
    \centering
    \includegraphics[width=8cm]{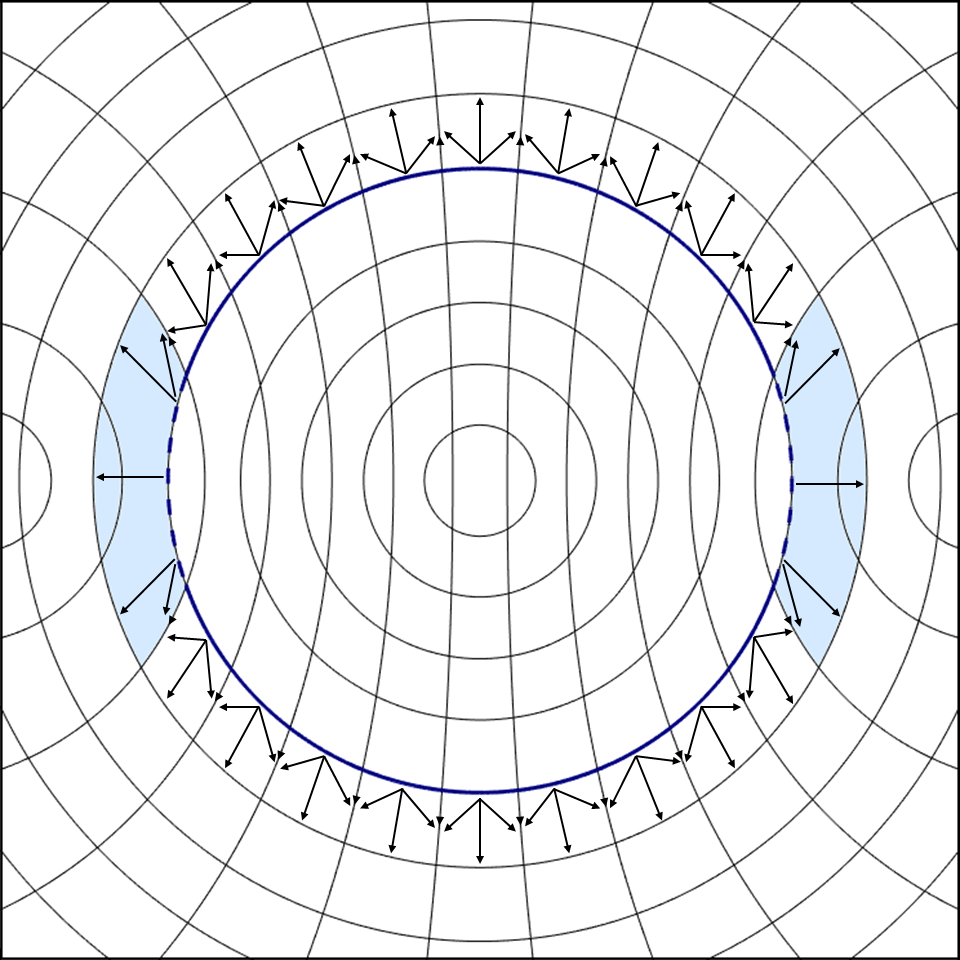}
    \caption{Stereographically projected spherical cross-section of the local unitary operator $S$, which sews the dark-blue toric code layer into the X-cube lattice, as used in the RG transformation for $S^2\times S^1$. $S$ is a product of CNOT gates corresponding to the arrows, which point from a control qubit to a target qubit. The arrows on the edges indicate gates that act on the edges oriented into (and out of) the plane and located at the adjacent vertices.
    Most of the CNOT gates are acting within cubes (depicated as curved sqaures above);
    within these cubes $S$ is the same as in \sfigref{fig:renormalization}{a}.
    The toric code plaquette operators extending out of the plane from the dashed blue lines are mapped to composite 3-cell operators on the 3-cells extending out of the plane from the shaded light-blue region.}
    \label{fig:S2S1}
\end{figure}

In this appendix we present examples of explicit transformations that add layers of toric code states to the X-cube model defined on $S^2\times S^1$ and $S^3$. In \figref{fig:S2S1} we depict a unitary transformation that sews a toroidal layer into the cellulation of $S^2\times S^1$ (\secref{sec:S2S1}).

\begin{figure*}
    \includegraphics[width=.23\textwidth]{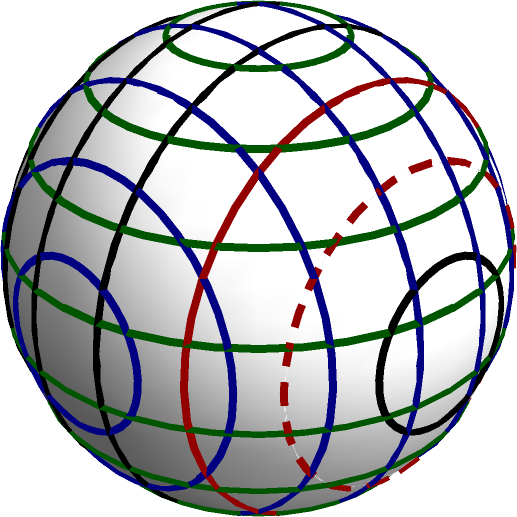}\hspace{.02\textwidth}
    \includegraphics[width=.35\textwidth]{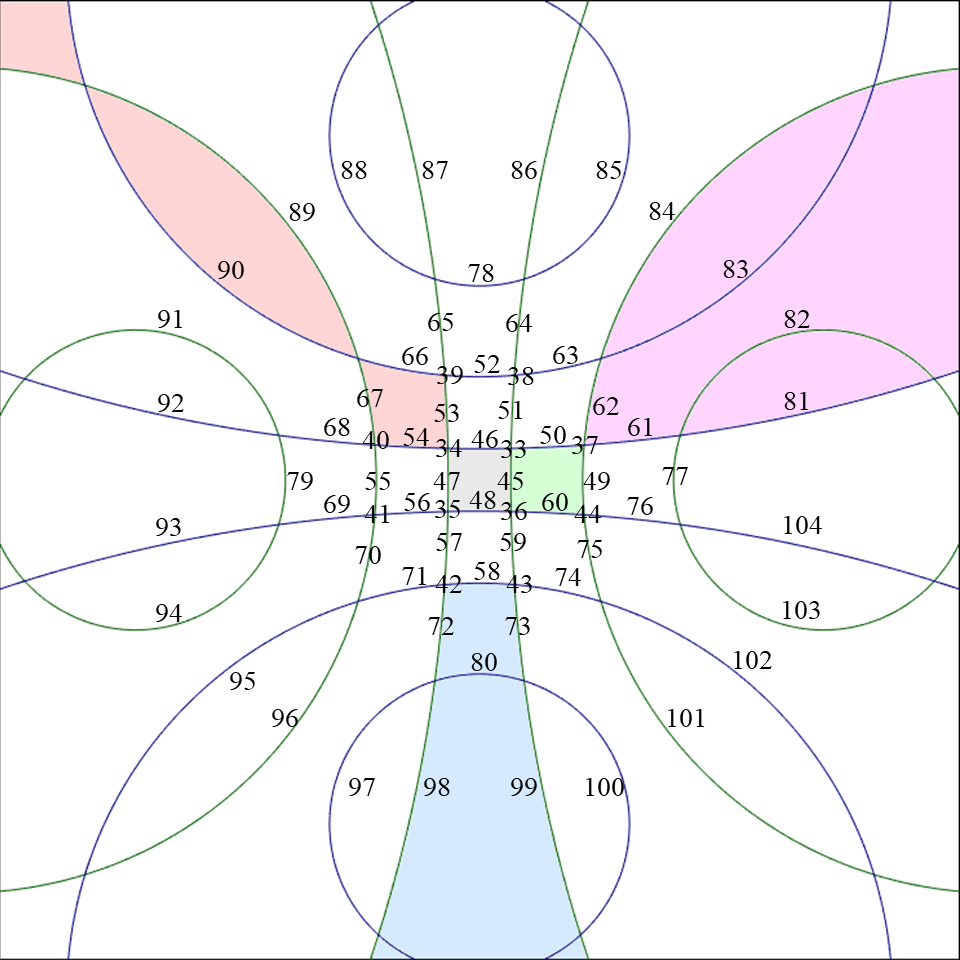}\hspace{.02\textwidth}
    \includegraphics[width=.35\textwidth]{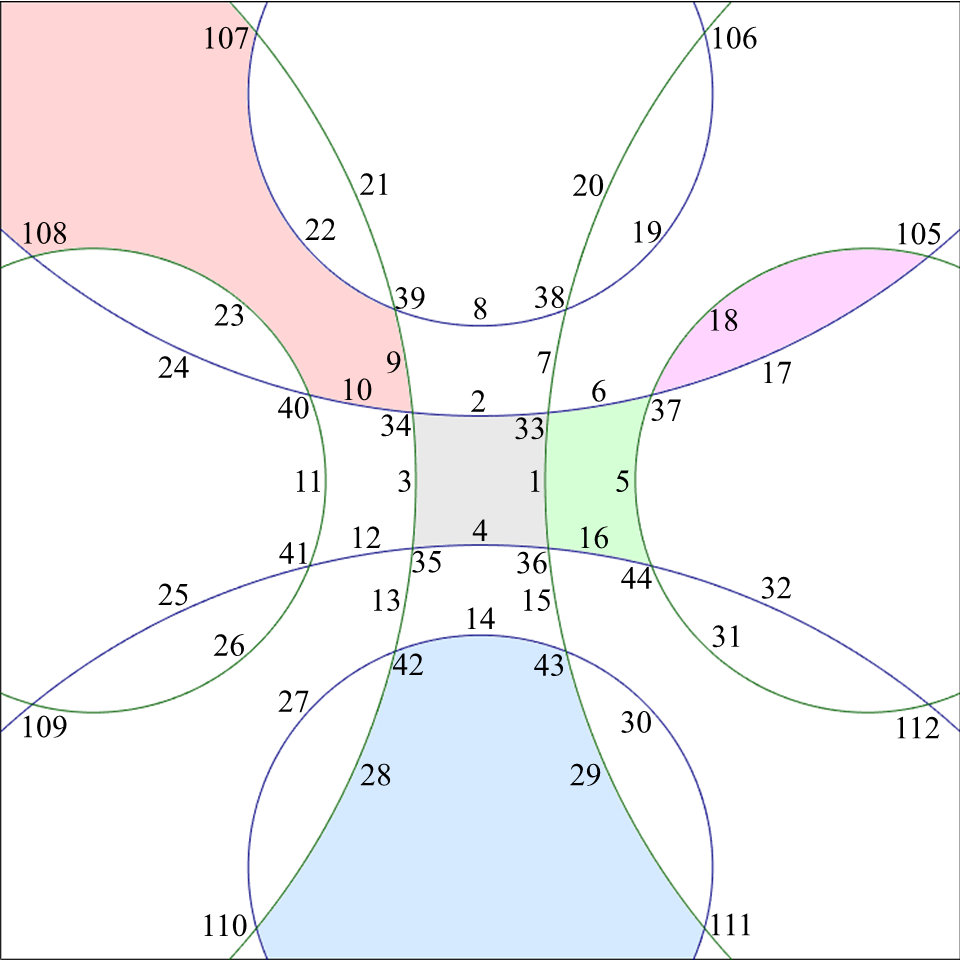} \\
    \begin{minipage}{.23\textwidth}
    {\bf (a)}
    \end{minipage}\hspace{.02\textwidth}
    \begin{minipage}{.35\textwidth}
    {\bf (b)}
    \end{minipage}\hspace{.02\textwidth}
    \begin{minipage}{.35\textwidth}
    {\bf (c)}
    \end{minipage}
    \caption{
    {\bf (a)} The equatorial cross-section of $S^3$ from \sfigref{fig:manifolds}{b}. We emphasize that the sphere drawn in (a) is not a leaf; it is merely a convenient cross-section.
    {\bf (b-c)} Stereographic projections of spherical leaves embedded in $S^3$.
    (b) and (c) intersect the equator (i.e. the spherical cross-section shown in (a)) at the solid and dashed red lines in (a), respectively.
    The green and blue lines represent links of the cellulation lying along the respective red-colored leaves in (a). The numbered vertices correspond to links that connect the two leaves, and thus share an index in (b) and (c). The unitary $S$ sews a toric code state (on the dashed red leaf) into the X-cube model on $S^3$. $S$ consists of a product of CNOT gates, and maps toric code plaquette operators on the dashed red layer to composite 3-cell operators lying between the two leaves. Some of the faces of these composite operators have been shaded in the figures as example. Plaquettes in (b) and (c) with corresponding colors indicate faces that belong to the same composite 3-cell.
    }
    
    \label{fig:S3}
\end{figure*}

In \figref{fig:S3} we illustrate the unitary transformation $S$ that sews a spherical layer into the cellulation of $S^3$ (\secref{sec:S3}).
Below we list all of the gates comprising $S$ that act on the qubits shown in \sfigref{fig:S3}{b-c}:
$\mathrm{CNOT}_{1,45}$, $\mathrm{CNOT}_{2,46}$, $\mathrm{CNOT}_{3,47}$, $\mathrm{CNOT}_{4,48}$, $\mathrm{CNOT}_{5,49}$, $\mathrm{CNOT}_{6,50}$, 
$\mathrm{CNOT}_{7,51}$, $\mathrm{CNOT}_{8,52}$, $\mathrm{CNOT}_{9,53}$, $\mathrm{CNOT}_{10,54}$, $\mathrm{CNOT}_{11,55}$, $\mathrm{CNOT}_{12,56}$, 
$\mathrm{CNOT}_{13,57}$, $\mathrm{CNOT}_{14,58}$, $\mathrm{CNOT}_{15,59}$, $\mathrm{CNOT}_{16,60}$, $\mathrm{CNOT}_{17,61}$, $\mathrm{CNOT}_{17,81}$, 
$\mathrm{CNOT}_{18,62}$, $\mathrm{CNOT}_{18,84}$, $\mathrm{CNOT}_{19,63}$, $\mathrm{CNOT}_{19,83}$, $\mathrm{CNOT}_{20,64}$, $\mathrm{CNOT}_{20,86}$, 
$\mathrm{CNOT}_{21,65}$, $\mathrm{CNOT}_{21,87}$, $\mathrm{CNOT}_{22,66}$, $\mathrm{CNOT}_{22,90}$, $\mathrm{CNOT}_{23,67}$, $\mathrm{CNOT}_{23,89}$, $\mathrm{CNOT}_{24,68}$, $\mathrm{CNOT}_{24,92}$, $\mathrm{CNOT}_{25,69}$, $\mathrm{CNOT}_{25,93}$, $\mathrm{CNOT}_{26,70}$, $\mathrm{CNOT}_{26,96}$, 
$\mathrm{CNOT}_{27,71}$, $\mathrm{CNOT}_{27,95}$, $\mathrm{CNOT}_{28,72}$, $\mathrm{CNOT}_{28,98}$, $\mathrm{CNOT}_{29,73}$, $\mathrm{CNOT}_{29,99}$, $\mathrm{CNOT}_{30,74}$, $\mathrm{CNOT}_{30,102}$,
$\mathrm{CNOT}_{31,75}$, $\mathrm{CNOT}_{31,101}$, $\mathrm{CNOT}_{32,76}$, $\mathrm{CNOT}_{32,104}$,
$\mathrm{CNOT}_{1,33}$,
$\mathrm{CNOT}_{1,36}$,
$\mathrm{CNOT}_{3,34}$,
$\mathrm{CNOT}_{3,35}$,
$\mathrm{CNOT}_{5,37}$,
$\mathrm{CNOT}_{5,44}$,
$\mathrm{CNOT}_{7,33}$,
$\mathrm{CNOT}_{7,38}$,
$\mathrm{CNOT}_{9,34}$,
$\mathrm{CNOT}_{9,39}$,
$\mathrm{CNOT}_{11,40}$,
$\mathrm{CNOT}_{11,41}$,
$\mathrm{CNOT}_{13,35}$,
$\mathrm{CNOT}_{13,42}$,
$\mathrm{CNOT}_{15,36}$,
$\mathrm{CNOT}_{15,43}$,
$\mathrm{CNOT}_{18,37}$,
$\mathrm{CNOT}_{18,105}$,
$\mathrm{CNOT}_{20,38}$,
$\mathrm{CNOT}_{20,106}$,
$\mathrm{CNOT}_{21,39}$,
$\mathrm{CNOT}_{21,107}$,
$\mathrm{CNOT}_{23,40}$,
$\mathrm{CNOT}_{23,108}$,
$\mathrm{CNOT}_{26,41}$,
$\mathrm{CNOT}_{26,109}$,
$\mathrm{CNOT}_{28,42}$,
$\mathrm{CNOT}_{28,110}$,
$\mathrm{CNOT}_{29,43}$,
$\mathrm{CNOT}_{29,111}$,
$\mathrm{CNOT}_{31,44}$, and 
$\mathrm{CNOT}_{31,112}$.


\begin{thebibliography}{50}
\bibitem [{foo()}]{foot:Betti}%
  \BibitemOpen
  \href@noop {} {}\bibinfo {note} {The $n^\text{th}$ Betti number with
  $\mathbb{Z}_p$ coefficients of a manifold $M$ is defined to be the rank of
  the $n^\text{th}$ homology group $H_n(M,\mathbb{Z}_p)$ over $\mathbb{Z}_p$.
  For the case of connected orientable surfaces, the first Betti number with
  $\mathbb{Z}_p$ coefficients is equal to twice the genus.}\BibitemShut {Stop}%
\end{thebibliography}
\end{document}